\documentstyle[12pt]{article}
\title{Towards a Unified Description of the Four Interactions in Terms of
Dirac-Bergmann Observables.}
\author{\large Luca Lusanna \\[3mm]
\em Sezione INFN di Firenze, \\
\em Largo E.Fermi 2, 50125 Firenze, Italy\\
\em E-mail; lusanna@fi.infn.it}

\begin{document}
\maketitle

Invited contribution to the book of the Indian National Science Academy for the
International Mathematics Year 2000 AD.

\begin{abstract}

A review is given of the status and developments of the research program
aiming to reformulate the physics of the four interactions at the classical 
level in a unified way in terms of Dirac-Bergmann observables with special
emphasis on the open mathematical, physical and interpretational problems.

\end{abstract}

At the classical level the accepted mathematical description of the four
interactions at the basis of our understanding of nature (gravitational, 
electromagnetic, weak and strong; without or with the not yet experimentally 
verified supersymmetry between half-integer and integer spin fields, i.e. 
between fermions and bosons) , is based on action principles which, due to 
manifest Lorentz invariance, to local gauge invariance (minimal coupling) 
and/or diffeomorphism invariances, make use of singular Lagrangians 
implying the Dirac-Bergmann theory of constraints\cite{dirac,bergm} for their
Hamiltonian formulation. While behind the gauge freedom of gauge theories 
proper there 
are Lie groups acting on some internal space so that the measurable quantities 
must be gauge invariant, the gauge freedom of theories invariant under
diffeomorphism groups of the underlying spacetime (general relativity, string
theory and reparametrization invariant systems of relativistic particles) 
concerns the arbitrariness for the observer in the choice of the definition
of ``what is space and/or time" (and relative times in the case of particles),
i.e. of the definitory properties either of spacetime itself or of the
measuring apparatuses.
This is the classical mathematical background on 
which our understanding of the quantum field theory of electromagnetic, weak 
and strong interactions in the modern BRS formulation is based. The same is true
for our attempts to build quantum gravity notwithstanding our actual incapacity 
to reconcile the influence of gravitational physics on the existence and
formulation of spacetime concepts with the basic ideas of quantum theory, which
requires a given absolute background spacetime.

Current research on electromagnetic, weak and strong interactions in special
relativity, namely in Minkowski spacetime, has partially bypassed the problem 
by the covariant approach based on the BRS symmetry which, at least at the 
level of the algebra of infinitesimal gauge transformations, allows a 
regularization and renormalization of the relevant theories inside the 
framework of local quantum field theory (see for instance Ref.\cite{weinberg}).
However, problems like the understanding of finite gauge transformations and
of the associated moduli spaces, the Gribov ambiguity dependence on the choice
of the function space for the fields and the gauge transformations, 
the confinement of quarks, the definition
of relativistic bound states and how to put them among the asymptotic states,
the nonlocality of charged states in quantum electrodynamics, not to speak of 
the foundational and practical problems posed by gravity, suggest that we 
should revisit the foundations of our theories.
It is not yet known  whether we can understand
which are the physical degrees of freedom hidden behind manifest gauge and/or
general covariance and whether we can firstly meaningfully reformulate classical
physics in terms of them and secondly to quantize the resulting theories. This
will require to abandon local field theory at the nonperturbative level and to
understand how to regularize and renormalize the Coulomb gauge of
electrodynamics to start with. Moreover, the special relativistic theories will
have to be reformulated in such a way to allow a natural transition to the 
coupling to gravity. Even if usually gravitational contributions are ignored
because they are too weak with respect to the other interactions, the existing
solution to the ultraviolet divergences of quantum field theory is 
distributional, so that, at least at the mathematical level, it is not 
justified to ignore gravity with all its
nonlinearities. In turn general relativity must be formulated in a way
allowing its deparametrization to recover physics in Minkowski spacetime when
the Newton constant is put equal to zero. One also needs a formulation in
which some notion of elementary particle exists so to recover Wigner's
definition based on the irreducible representations of the Poincar\'e group
in Minkowski spacetime with the further enrichment of the known good quantum 
numbers for their classification. Moreover, one needs some way out from the
``problem of time"\cite{ish,kuchar1,butter}, since neither any consistent way 
to quantize time (is it a necessity?), and generically any timelike variable,
nor a control on the associated problem of the relative times of a system of 
relativistic particles are known. Finally, one has to find a solution to the 
more basic problem of how to identify physically spacetime points in
Einstein's formulation of general relativity, where general covariance deprives
the mathematical points of the underlying 4-manifold of any physical reality
\cite{rove,rov},
while, on the experimental side (space physics, gravitational waves detectors)
, we are employing a theory of measurements of proper times and spacelike 
lengths which presuppones the individuation of points. This problem will appear
also in the nowaday most popular program of unification of all the interactions 
in a supersymmetric way, i.e. in superstring theory and in its searched 
M-theory extension (see for instance Ref.\cite{polc}; string theory will not be
touched in this review), when someone will be able to reformulate it in a
background independent way.

These motivations induced me to revisit the classical Hamiltonian formulation
of theories described by singular Lagrangians trying to choose the mathematical
frameworks which at each step looked more natural to clarify the physical
interpretational problems by means of the use of suitable adapted coordinates.
In particular, after many years of dominance of the point of view privileging
manifest Lorentz, gauge and general covariance at the price of loosing control
on the physical degrees of freedom and on their deterministic evolution (felt
as a not necessary luxury only source of difficulties and complications), I went 
back to the old concept of Dirac observables, namely of those gauge invariant
deterministic variables which describe a canonical basis of measurable
quantities for the electromagnetic, weak and strong interactions in
Minkowski spacetime. Instead, in general relativity, due to the problem of the
individuation of the points of spacetime, measurable quantities have a more 
complex identification, which coincides with Dirac's observables (in any case
indispensable for the treatment of the Cauchy problem) only in a 
completely fixed gauge (total breaking of general covariance).

In the next Sections I will review the various achievements of the
program at the present stage of development (see Refs.\cite{re} for previous
reviews). Since there is too vast a bibliography to be covered in this review,
I made the choice to concentrate it on my point of view omitting to quote
many aspects of the theory and the work of many researchers.

\section{Singular Lagrangians, Presymplectic Geometry, the Shanmugadhasan
Canonical Transformations and Generalized Coulomb Gauges in Minkowski
Spacetime.}

A) If a finite-dimensional system with configuration space $Q$ [$q^i$, i=1,..,N,
are local coordinates in a global (assumed to exist for the sake of simplicity)
chart of the atlas of Q; $(t,q^i(t))$ is a point in $R\times Q$, where $R$ is 
the time axis; ${\dot q}^i(t)=dq^i(t)/dt$] is described by a singular
Lagrangian $L$ [so that the Hessian matrix is degenerate: $det\, \Big( 
\partial^2L/\partial {\dot q}^i\partial {\dot q}^j\Big) =0$], its 
Euler-Lagrange equations are in general a mixture of equations i) depending
only on the $q^i$ (holonomic constraints); ii) depending only on $q^i$ and
${\dot q}^i$ (Lagrangian, in general nonholonomic, constraints and/or intrinsic 
first order equations of motion violating the so called second order 
differential equation (SODE) conditions); iii) depending on $q^i$, ${\dot q}^i$,
${\ddot q}^i$ (genuine second order equations of motion, which however cannot
be put in normal form, i.e. solved in the ${\ddot q}^i$). More equations of the
types i) and ii) can be deduced from the Euler-Lagrange equations and their 
time derivatives. The study of this type of degenerate equations can be traced
back to Levi-Civita\cite{levi}.
The solutions of the Euler-Lagrange equations depend on 
arbitrary functions of time, namely they are not deterministic.

The canonical momenta $p_i=\partial L/\partial {\dot q}^i$ are not independent:
there are relations among them $\phi_{\alpha}(q,p)\approx 0$ called primary
Hamiltonian constraints, which define a submanifold $\gamma$ of the cotangent 
space $T^{*}Q$ [the model is defined only on this submanifold; one uses the
Poisson brackets of $T^{*}Q$ in a neighbourhood of $\gamma$ and Dirac's weak
equality $\approx$ means that the equality sign cannot be used inside Poisson 
brackets]. The canonical Hamiltonian $H_c(q,p)$ has to be replaced by the Dirac 
Hamiltonian $H_D=H_c+\sum_{\alpha} \lambda_{\alpha}(t)\phi_{\alpha}$, which 
knows the restriction to the submanifold $\gamma$ due to the arbitrary Dirac 
multipliers $\lambda_{\alpha}(t)$. The time constancy of the primary constraints
, $\partial_t \phi_{\alpha}=\{ \phi_{\alpha},H_D\} \approx 0$, either pruduces
secondary Hamiltonian constraints or determines some of the Dirac multipliers.
This procedure is repeated for the secondary constraints (this is the Dirac-
Bergmann algorithm) and so on. At the end there is a final set of constraints
$\chi_a\approx 0$ defining the final submanifold $\bar \gamma$ of $T^{*}Q$ on
which the dynamics is consistently restricted, and a final Dirac Hamiltonian
with a reduced set of arbitrary Dirac multipliers describing the remaining
indetermination of the time evolution. The constraints are divided into two
subgroups: i) the first class ones $\chi^{(1)}_m\approx 0$, having weakly
zero Poisson bracket with all constraints and being the generators of the gauge 
transformations of the theory (the associated vector fields $\{ ., \chi^{(1)}
_m\}$ are tangent to $\bar \gamma$); ii) the second class ones $\chi^{(2)}_n
\approx 0$ (their number is even) with $det\, \Big( \{ \chi^{(2)}_{n_1},
\chi^{(2)}_{n_2}\} \Big) \not= 0$, corresponding to pairs of inessential
eliminable variables (the associated vector fields are normal to $\bar \gamma$).
The solutions of the Hamilton-Dirac equations with the final Dirac Hamiltonian 
depend on as many arbitrary functions of time as the left Dirac multipliers.
The restriction of the symplectic 2-form of $T^{*}Q$ to $\bar \gamma$ is a 
closed degenerate 2-form, which in case of only first class constraints 
generates a so called presymplectic geometry: $\bar \gamma$ is said to be a
presymplectic manifold coisotropically embedded in $T^{*}Q$ [see 
Ref.\cite{lich,go} for what is known on presymplectic structures (they are dual
to Poisson structures, but much less studied not being connected with 
integrable systems) and on the more general ones when second class constraints 
are present]. 
When many mathematical conditions are satisfied, the vector fields associated 
with the first class constraints (they are in the kernel of the degenerate 
2-form on $\bar \gamma$) generate a foliation of the submanifold $\bar \gamma$:
each leaf (Hamiltonian gauge orbit) contains all the configurations which are 
gauge equivalent and which have to be considered as the same physical 
configuration\cite{dirac} (equivalence class of gauge equivalent 
configurations); the canonical Hamiltonian $H_c$ (if it is not $H_c\approx 0$)
generates an evolution which maps one leaf into the others.
 Therefore, the physical reduced phase space is obtained: i)
by eliminating as many pairs of conjugate variables as second class constraints
by means of the so called associated Dirac brackets; ii) by going to the
quotient with respect to the foliation (a representative of the reduced phase 
space can be build by adding as many gauge-fixing constraints as first class
ones, so to obtain a set of second class constraints). In general this
procedure breaks the original Lorentz invariance.

Let us remark that only the primary first class constraints are associated with 
arbitrary Dirac multipliers. The secondary, tertiary... first class constraints
are, in general, present in the canonical Hamiltonian $H_c$ multiplied by well
defined functions of $q^i$, ${\dot q}^i$, which turn out to be arbitrary 
because they are not determined by the Hamilton-Dirac equations (they are 
gauge variables). This contradicts the Dirac conjecture\cite{dirac} that the
secondary first class constraints can be added to the Dirac Hamiltonian with 
extra multipliers (the resulting extended Dirac Hamiltonian would not allow
the reconstruction of the original singular Lagrangian by inverse Legendre
transformation; since the difference in the dynamics is only off-shell, this
explains why the extended Hamiltonian is used in the BFV approach\cite{bfv}).
The natural way to add gauge-fixing constraints when there are secondary first 
class constraints\cite{sugano}, is to start giving the gauge fixings to the 
secondary constraints. The requirement of time constancy of these gauge fixings
will generate the gauge fixings for the primary first class constraints and the
time constancy of these new gauge fixings will determine the Dirac multipliers
eliminating every residual gauge freedom.

The Dirac observables
are the gauge invariant functions on the reduced phase space, on which there 
is a deterministic evolution generated by the projection of the canonical 
Hamiltonian. Therefore, the main problem is to find a (possibly global) Darboux 
coordinate chart of the reduced phase space, namely a canonical basis of Dirac 
observables (or at least a Poisson algebra of them, according to Ref.\cite{sny}
).

One would expect that when this is not
possible, the relativistic system is intrinsically ill defined already at
the classical level: at the quantum level this should manifest itself
with the presence of not curable anomalies (which can be present also for
a classically well defined system). Since the mathematical theory of the
anomalies relies on cohomological properties of the manifolds (like Q and
$\bar \gamma$) relevant to the description of the system, which have to
be defined already at the classical level, one expects that a classical
background of these properties in the form of obstructions to the determination
of the observables should be present in the
theory of classical gauge canonical transformations.

When there is reparametrization invariance of the original action
$S=\int dt L$, the canonical Hamiltonian vanishes and the reduced phase space
is said frozen (like it happens in Hamilton-Jacobi theory).
When the
canonical Hamiltonian vanishes, both kinematics and dynamics are contained
in the first class constraints describing the system: these can be
interpreted as generalized Hamilton-Jacobi equations\cite{domi}, so that
the Dirac observables turn out to be the Jacobi data. When there is a 
kinematical symmetry group, like the Galileo or Poincar\'e groups, an evolution
may be reintroduced by using the energy generator as Hamiltonian.

In a series of papers\cite{lu1,lu2,lu3,lu4,lu5,lu6} I made a reformulation of
the general theory of singular Lagrangians and Hamiltonian constraints based on
an extension of the second Noether theorem\cite{noether} to include also 
second class constraints. By means of the resulting Noether
identities the Dirac-Bergmann algorithm was reproduced at the Lagrangian level.
All the obscure and/or ambigous points of the theory were clarified. The 
understanding\cite{lu2} of the pathological examples known in the literature 
led to the discovery of third- and fourth-class constraints [with their
associated singularities of the Jacobi equations (linearization of the
Euler-Lagrange equations) and their connection with the
reject of the Dirac conjecture about adding the secondary first class 
constraints to the Dirac Hamiltonian with extra Dirac multipliers]
and of the phenomena of
proliferation of constraints, ramification and joining of chains of constraints.
Also the classification of all possible patterns of second class constraints 
was given\cite{lu6}. All these phenomena have their counterpart in the study
of the Euler-Lagrange equations for a singular Lagrangian in the second-order
formalism. In Ref.\cite{lu5} there is also the status of the art for the
much more difficult and still incomplete 
first-order formulation of the theory on the
tangent space $TQ$ or on the first jet bundle $J^1(Q)\approx TQ\times R$,
while in Ref.\cite{lu4} there is the connection with BRS theory.

B) Now I will delineate the main steps for the determination of the Dirac 
observables for the case in which only primary first class constraints $\phi
_{\alpha}\approx 0$ are present at the Hamiltonian level.

The Euler-Lagrange equations associated with a singular Lagrangian do not
determine the gauge part of the extremals. However it cannot be totally 
arbitrary, but must be compatible with the algebraic properties of the 
Noether gauge transformations induced by the first class constraints under 
which the action is either invariant or quasi-invariant as implied by the
second Noether theorem. In the Hamiltonian formulation these properties are
contained in the structure constants, or functions, of the Poisson brackets
of the first-class constraints among themselves [$\{ \phi_{\alpha}, \phi
_{\beta}\} = C_{\alpha\beta\gamma} \phi_{\gamma}$, $\{ \phi_{\alpha},H_c\} =
C_{\alpha\beta}\phi_{\beta}$]
and the gauge arbitrariness of the 
trajectories is described by the Dirac multipliers appearing in the Dirac
Hamiltonian. In both formulations one has to add extra equations, the
either Lagrangian or Hamiltonian multitemporal equations\cite{lu3}, to have a
consistent determination of the gauge part of the trajectory (see the
generalized Lie equations of Ref.\cite{bv}). These equations are obtained by
rewriting the variables $q^i(t)$, $p_i(t)$ in the form $q^i(t,\tau_{\alpha})$,
$p_i(t,\tau_{\alpha})$, and by assuming that the original t-evolution
generated by the Dirac Hamiltonian $H_D=H_c+\sum_{\alpha}\lambda_{\alpha}(t)
\phi_{\alpha}$ is replaced by: i) a deterministic t-evolution generated by
$H_c$; ii) a $\tau_{\alpha}$-evolution (reassorbing the arbitrary Dirac
multipliers $\lambda_{\alpha}(t)$), for each $\alpha$, generated in a
suitable way by the first class constraints $\phi_{\alpha}$. The
$\tau_{\alpha}$-dependence of $q^i$, $p_i$ determined by these multitemporal
(or better multiparametric) equations, which are integrable due to the 
first-class property of the constraints, describes their dependence on the 
gauge orbit containing the given Cauchy data for the Hamilton-Dirac equations.
From the point of view of the study of the multitemporal equations, the 
secondary first class constraints are treated like the primary ones, namely as 
if there would be associated extra Dirac multipliers, and one should use as
canonical Hamiltonian $H_c$ restricted to zero value of the secondary
constraints.

When the Poisson brackets of the
Hamiltonian first class constraints imply a canonical realization of a
Lie algebra, the extra Hamiltonian
multitemporal equations have the first class
constraints as Hamiltonians (so that the Dirac Hamiltonian is reduced to the
canonical Hamiltonian) and the time parameters (replacing the Dirac
multipliers) are the coordinates of a group manifold for a Lie group whose
algebra is the given Lie algebra: they enter in the multitemporal
equations via a set of left invariant vector fields $Y_{\alpha}$ on the group 
manifold [$Y_{\alpha} A(q,p)=\{ A(q,p), \phi_{\alpha}\}$]. In the ideal case
in which the gauge foliation of $\bar \gamma$
is nice, all the leaves (or gauge orbits) are
diffeomorphic and in the simplest case all of them are diffeomorphic to the
group manifold of a Lie group. In this ideal case to rebuild a gauge
orbit from one of its points (and therefore to determine the gauge part
of the trajectories passing through that point) one needs the Lie equations
associated with the given Lie group: the Hamiltonian multitemporal equations
are generalized Lie equations describing all the gauge orbits simultaneously.
In a generic case this description holds only locally for a set of
diffeomorphic orbits, also in the case of systems invariant under
diffeomorphisms.

Once one has solved the multitemporal equations, the next step is the 
determination of a  Shanmugadhasan canonical transformation\cite{sha}. In the 
finite dimensional case general theorems\cite{kulk} connected with the Lie 
theory of function groups\cite{lie} ensure the existence of local canonical 
transformations from the original canonical variables $q^i$, $p_i$, in terms 
of which the first class constraints (assumed globally defined) have the form 
$\phi_{\alpha}(q,p)\approx 0$, to canonical bases $P_{\alpha}$, $Q_{\alpha}$, 
$P_A$, $Q_A$, such that the equations $P_{\alpha}\approx 0$ locally define the 
same original constraint manifold (the $P_{\alpha}$ are an Abelianization of the
first class constraints); the $Q_{\alpha}$ are the adapted Abelian gauge
variables describing the gauge orbits (they are a realization of the times
$\tau_{\alpha}$ of the multitemporal equations in terms of variables $q^i$,
$p_i$); the $Q_A$, $P_A$ are an adapted canonical basis of Dirac
observables. These canonical transformations are the basis of the Hamiltonian
definition of the Faddeev-Popov measure of the path integral\cite{fp} and
give a trivialization of the BRS construction of observables (the BRS method
works when the first class constraints may be Abelianized\cite{hennea}).
Therefore the problem of the search of the Dirac observables
becomes the problem of finding Shanmugadhasan canonical transformations. The
strategy is to find abelianizations $P_{\alpha}$ of the original constraints,
to solve the multitemporal equations for $q^i$, $p_i$ associated with the
$P_{\alpha}$, to determine the multitimes $Q_{\alpha}=\tau_{\alpha}$ and to 
identify the
Dirac observables $P_A$, $Q_A$ from the remaining original variables, i.e.
from those their combinations independent from $P_{\alpha}$ and $Q_{\alpha}$.
Second class constraints, when present, are also taken into account by the
Shanmugadhasan canonical transformation\cite{sha}.

Putting 
equal to zero the Abelianized gauge variables one defines a local gauge of the
model. If a system with constraints admits one (or more) global
Shanmugadhasan canonical transformations, one obtains one (or more) privileged 
global gauges in which the physical Dirac observables are globally defined and
globally separated from the gauge degrees of freedom [for
systems with a compact configuration space Q this is impossible]. These
privileged gauges (when they exist) can be called generalized Coulomb gauges.
When the system under investigation has some
global symmetry group, the associated theory of the momentum map\cite{souriau}
is a source of globality.

C)
Now all the physical systems defined in the flat Minkowski spacetime, have the 
global Poincare' symmetry. This suggests to study the structure of the 
constraint manifold $\bar \gamma$ from the point of view of the
orbits of the Poincare' group. If $p^{\mu}$ is the total momentum of the
system, the constraint manifold has to be divided in four strata (some of
them may be absent for certain systems) according to whether $p^2 > 0$,
$p^2=0$, $p^2 < 0$ or $p^{\mu}=0$. Due to the different little groups of the
various Poincare' orbits, the gauge orbits of different sectors will not be
diffeomorphic. Therefore the manifold $\bar \gamma$ is a stratified manifold
and the gauge foliations of relativistic systems are
nearly never nice, but rather one has to do with singular foliations.

For an acceptable relativistic system the stratum $p^2 < 0$ has to be absent
to avoid tachyons. To study the strata $p^2=0$ and $p^{\mu}=0$ one has to add
these relations as extra constraints. For all the strata the next step (see
however the next Section) is to do a canonical transformation from the original 
variables to a new set consisting of center-of-mass variables $x^{\mu}$, 
$p^{\mu}$ and of variables relative to the center of mass. Let us now consider 
the stratum $p^2 > 0$. By using the standard Wigner boost $L^{\mu}_{\nu}(p,
{\buildrel \circ \over p})$  ($p^{\mu}=L^{\mu}_{\nu}(p,{\buildrel \circ \over
p}){\buildrel \circ \over p}^{\nu}$, ${\buildrel \circ \over p}^{\mu}=\eta 
\sqrt {p^2} (1;\vec 0 )$, $\eta = sign\, p^o$), one boosts the relative 
variables at rest.
The new variables are still canonical and the base is completed by $p^{\mu}$
and by a new center-of-mass coordinate ${\tilde x}^{\mu}$, differing from 
$x^{\mu}$ for spin terms. The variable ${\tilde x}^{\mu}$ has complicated 
covariance properties; instead the new relative variables are either Poincare'
scalars or Wigner spin-1 vectors, transforming under the group O(3)(p) of
the Wigner rotations induced by the Lorentz transformations. A final
canonical transformation\cite{longhi}, leaving fixed the relative variables, 
sends the center-of-mass coordinates ${\tilde x}^{\mu}$, $p^{\mu}$ in the new 
set $p\cdot {\tilde x}/\eta \sqrt {p^2}=p\cdot x/\eta \sqrt {p^2}$ (the time
in the rest frame), $\eta \sqrt {p^2}$ (the total mass), $\vec k =\vec p
/\eta \sqrt {p^2}$ (the spatial components of the 4-velocity $k^{\mu}=
p^{\mu}/\eta \sqrt {p^2}$, $k^2=1$), $\vec z=\eta \sqrt {p^2}( {\vec {\tilde
x}}-{\tilde x}^o\vec p/p^o)$. $\vec z$ is a noncovariant center-of-mass 
canonical 3-coordinate  multiplied by the total mass: it is the classical 
analog of the Newton-Wigner position operator (like it, $\vec z$ is covariant 
only under the little group O(3)(p) 
of the timelike Poincar\'e orbits). Analoguous 
considerations could be done for the other sectors. In Ref.\cite{spin} there
is the definition of other canonical bases, the spin bases, adapted to the 
spin Casimir of the Poincar\'e group.

The nature of the relative variables depends on the system. The first class
constraints, once rewritten in terms of the new variables, can be manipulated
to find suitable global and Lorentz scalar Abelianizations.
Usually there is a combination of the constraints which determines $\eta
\sqrt {p^2}$, i.e. the mass spectrum, so that the time in the rest frame
$p\cdot x/\eta \sqrt {p^2}$ is the conjugated Lorentz scalar gauge variable.
The other constraints eliminate some of the relative variables (in particular
the relative energies for systems of interacting relativistic particles and the
string): their conjugated coordinates (the relative times) are the other gauge 
variables: they are identified with a possible set of time parameters by the 
multitemporal equations. The Dirac observables (apart from the center-of-mass
ones $\vec k$ and $\vec z$) have to be extracted from the
remaining relative variables and the construction shows that they will be
either Poincare' scalars or Wigner covariant objects.
In this way in each stratum preferred global Shanmugadhasan canonical
transformations are identified, when no other kind of obstruction to
globality is present inside the various strata.

D)
In gauge field theories the situation is more complicated, becouse the
theorems ensuring the existence of the Shanmugadhasan canonical
transformation have not been extended to the infinite-dimensional case.
One of the reasons is that some of the constraints can now be interpreted as
elliptic equations and they can have zero modes. Let us consider the stratum
$p^2 > 0$ of free Yang-Mills theory as a prototype and its first class
constraints, given by the Gauss laws and by the vanishing of the time
components of the canonical momenta. The problem of the zero modes will
appear as a singularity structure of the gauge foliation of the allowed
strata, in particular of the stratum $p^2 > 0$. This phenomenon was discovered
in Ref.\cite{cone} by studying the space of solutions of Yang-Mills and Einstein
equations, which can be mapped onto the constraint manifold of these theories 
in their Hamiltonian description. It turns out that the space of solutions
has a "cone over cone" structure of singularities: if we have a line of 
solutions with a certain number of symmetries, in each point of this line
there is a cone of solutions with one less symmetry. In the Yang-Mills case
the ``gauge symmetries" of a gauge potential are connected with the generators 
of its stability group, i.e. with the subgroup of those special gauge 
transformations which leave invariant that gauge potential (this is the
Gribov ambiguity for gauge potentials; there is also a more general Gribov
ambiguity for field strengths, the ``gauge copies" problem).
Since the Gauss laws are the
generators of the gauge transformations (and depend on the chosen gauge 
potential through the covariant derivative), this means that for a gauge
potential with non trivial stability group those combinations of the
Gauss laws corresponding to the generators of the stability group cannot
be any more first class constraints, since they do not generate effective
gauge transformations but special symmetry transformations. This problematics
has still to be clarified, but it seems that in this case these components
of the Gauss laws become third class constraints, which are not generators
of true gauge transformations. This new kind of constraints was introduced
in Refs.\cite{lu2,lu5} in the finite dimensional case as a
result of the study of some examples, in which the Jacobi equations (the
linearization of the Euler-Lagrange equations) are singular, i.e. some
of their solutions are not infinitesimal deviations between two
neighbouring extremals of the Euler-Lagrange equations. This interpretation
seems to be confirmed by the fact that the singularity structure discovered
in Ref.\cite{cone} follows  from the existence of singularities of the 
linearized Yang-Mills and Einstein equations. These problems are part of the 
Gribov ambiguity, which, as a consequence, induces an extremely complicated 
stratification and also singularities in each Poincar\'e stratum of $\bar 
\gamma$.

Other possible sources of singularities of the gauge foliation of Yang-Mills
theory in the stratum $p^2 > 0$ may be: i) different classes of gauge
potentials identified by different values of the field invariants; 
ii) the orbit structure of the
rest frame (or Thomas) spin $\vec S$, identified by the Pauli-Lubanski Casimir
$W^2=-p^2{\vec S}^2$ of the Poincare' group.

The final outcome of this structure of singularities is that the reduced
phase-space, i.e. the space of the gauge orbits, is in general a 
stratified manifold with singularities\cite{sny}. In the stratum $p^2 > 0$ of
the Yang-Mills theory these singularities survive the Wick rotation to the
Euclidean formulation and it is not clear how the ordinary path integral
approach and the associated BRS method can take them into account. The search
of a global canonical basis of Dirac observables for each stratum of the
space of the gauge orbits can give a definition of the measure of the
phase space path integral, but at the price of a non polynomial
Hamiltonian. Therefore, if it is not possible to eliminate the Gribov
ambiguity (assuming that it is only a mathematical obstruction without any 
hidden physics), the existence of global Dirac observables for Yang-Mills theory
is very problematic.

E)
Firstly, inspired by Ref.\cite{dira} where a canonical basis of Dirac 
observables was found for the electromagnetic field interacting with a fermion
field (whose Dirac observable is a fermion field dressed with a Coulomb cloud),
the canonical reduction to noncovariant 
generalized Coulomb gauges, with the determination of the physical Hamiltonian
as a function of a canonical basis of Dirac's observables, has been achieved for
the following isolated systems (for them one asks that the 10 conserved 
generators of the Poincar\'e algebra are finite so to be able to use group 
theory; theories with external fields can only be recovered as limits in some
parameter of a subsystem of the isolated system): 

1) Relativistic particle mechanics. Its importance
stems from the fact that quantum field theory has no
particle interpretation: this is forced on it by means of the asymptotic states
of the reduction formalism
which correspond to the quantization of independent one-body systems
described by relativistic mechanics [or relativistic pseudoclassical mechanics
\cite{l1}, when one adds Grassmann variables to describe the intrinsic spin].
Besides the scalar particle ($p^2-m^2\approx 0$ or $p^2\approx 0$), one has
control on: i) the pseudoclassical electron\cite{l2} ($p_{\mu}\xi^{\mu}-m\xi_5
\approx 0$ or $p_{\mu}\xi^{\mu}\approx 0$, where $\xi^{\mu}, \xi_5$ are 
Grassmann variables; $p^2-m^2\approx 0$ or $p^2\approx 0$ are implied; after
quantization the Dirac equation is reproduced); 
ii) the pseudoclassical neutrino\cite{l3} ($p_{\mu}\xi^{\mu}+{i\over 3}\epsilon
^{\mu\nu\rho\sigma}p_{\mu}\xi_{\nu}\xi_{\rho}\xi_{\sigma}\approx 0$, $p^2
\approx 0$, giving the Weyl particle wave equation $p_{\mu}\gamma^{\mu}(1-
\gamma_5)\psi(x)=0$ after quantization); iii) the
pseudoclassical photon\cite{l4} ($p^2\approx 0$, $p_{\mu}\theta^{\mu}\approx 0$,
$p_{\mu}\theta^{{*}\mu}\approx 0$, $\theta^{*}_{\mu}\theta^{\mu}\approx 0$,
where $\theta^{\mu}, \theta^{{*}\mu}$ are a pair of complex Grassmann 
four-vectors 
to describe helicity $\pm 1$; after quantization one obtains the photon
wave equations $\bar \sqcup A^{\mu}(x)=0$, $\partial_{\mu}A^{\mu}(x)=0$;
the Berezin-Marinov Grassmann distribution function allows to recover the
classical polarization matrix of classical light and, in quantization, the
quantum polarization matrix with the Stokes parameters); 
iv) the vector particle or pseudoclassical massive photon\cite{l5} [$p^2-\mu^2+
(1-\lambda)p_{\mu}\theta^{{*}\mu}p_{\nu}\theta^{\nu}\approx 0$,
$\theta^{*}_{\mu}\theta^{\mu}\approx 0$, which, after quantization, reproduce
the Proca-like wave equation $(\bar \sqcup +\mu^2)A^{\mu}(x)-(1-\lambda )
\partial^{\mu}\partial_{\nu}A^{\nu}(x)=0$].

The most important two-body system is the DrozVincent-Todorov-Komar model
\cite{droz} with an arbitrary action-at-a-distance interaction instantaneous
in the rest frame as shown by its energy-momentum tensor\cite{l6} [$p_i^2-m_i^2+
V(r^2_{\perp})\approx 0$, i=1,2, $r^{\mu}_{\perp}=(\eta^{\mu\nu}-p^{\mu}p^{\nu}
/p^2)r_{\nu}$, $r^{\mu}=x_1^{\mu}-x^{\mu}_2$, $p_{\mu}=p_{1\mu}+p_{2\mu}$]. This
model has been completely understood both at the classical and quantum level
\cite{longhi}. Its study led to the identification of 
a class of canonical transformations (utilizing the standard Wigner boost for
timelike Poincar\'e orbits) which allowed to understand how to define suitable
center-of-mass and relative variables (in particular a suitable relative energy
is determined by a combination of the two first class constraints, so that the 
relative time variable is a gauge variable), how to find a quasi-Shanmugadhasan 
canonical transformation adapted to the constraint determining the relative
energy, how to separate the four, topologically disjoined, branches of the mass 
spectrum (it is determined by the other independent combination of the 
constraints; therefore, there is a distinct Shanmugadhasan canonical 
transformation for each branch). At the quantum level it was possible to find 
four physical scalar products, compatible with both the resulting coupled
wave equations (i.e. independent from the relative and the absolute rest-frame 
times): they have been found as generalization of the two existing scalar
products of the Klein-Gordon equation: all of them are non-local even in the
limiting free case and differ among themselves for the sign of the norm of
states on different mass-branches. This example shows that the physical scalar
product knows the functional form of the constraints.

The connection with the Bethe-Salpeter equation of the quantized model has been
studied in Ref.\cite{saz}, where it is shown that the constraint wave function 
can be obtained from the Bethe-Salpeter one by multiplication for a delta 
function containing the relative energy  to exclude its spurious solutions
(non physical excitations in the relative energy).
The extension of the model to two pseudoclassical electrons and to an electron
and a scalar has been done in Ref.\cite{crater}, and the first was used to get 
good fits to meson spectra.

The previous canonical transformations were then extended to N free particles
described by N mass-shell first class constraints $p^2_i-m^2_i\approx 0$
\cite{l7}: N-1 suitable relative energies are determined by N-1 combinations of
the constraints (so that the conjugate N-1 relative times are gauge variables),
while the remaining combination determines the $2^N$ branches of the mass
spectrum. The N gauge freedoms associated with these N combinations of the
first class constraints are the freedom of the observer: i) in the choice of the
time parameter to be used for the overall evolution of the isolated system;
ii) in the choice of the description of the relative motions with any given 
delay among the pairs of particles.

In Ref.\cite{l8} 2- and N-body Newton mechanics was reformulated in a
multitemporal way in terms of N first class constraints obtained from the 
relativistic ones in the limit $c \rightarrow \infty$. After a comparison with 
predictive mechanics, it was shown that the ``no-interaction-theorem" (namely
that the multitemporal configurational and canonical position coordinates of a
particle coincide only in absence of interactions) exists also at the 
nonrelativistic level, being a property of the multitemporal description of 
particles and not of the kinematical symmetry group.

2) 
Both the open and closed Nambu string, after an initial study with light-cone
coordinates, have been treated\cite{colomo} 
along the lines of the two-body model in the stratum $p^2 > 0$. Both Abelian 
Lorentz scalar constraints and gauge variables
have been found and globally decoupled, and a redundant set of Dirac's
observables $[\vec z,\vec k,{\vec {\tilde a}}_n]$ has been found. It remains an
open problem whether one can extract a global canonical basis of Dirac's
observables from the Wigner spin 1 vectors ${\vec {\tilde a}}_n$, which satisfy
sigma-model-like constraints; if this basis exists, it would define the
Liouville integrability of the Nambu string and would clarify whether there is
any way to quantize it in four dimensions.

3) Yang-Mills theory with Grassmann-valued
fermion fields \cite{lusa} in the case of a trivial principal
bundle over a fixed-$x^o$ $R^3$ slice of Minkowski spacetime with suitable
Hamiltonian-oriented boundary conditions; this excludes monopole solutions 
(to have them, even if they have been not yet found experimentally,
one needs a nontrivial bundle and a variational principle
formulated on the bundle\cite{marmo}, because the gauge potentials on
Minkowski spacetime are not globally defined) and,
since $R^3$ is not compactified, one has only winding number and no instanton
number. After a discussion of the
Hamiltonian formulation of Yang-Mills theory, of its group of gauge
transformations and of the Gribov ambiguity, the theory has been studied in
suitable  weighted Sobolev spaces where the Gribov ambiguity is absent
\cite{vince} and the global color charges are well defined.
The global Dirac observables are the transverse quantities ${\vec A}_{a\perp}
(\vec x,x^o)$, ${\vec E}_{a\perp}(\vec x,x^o)$ and fermion fields dressed
with Yang-Mills (gluonic) clouds. The nonlocal and nonpolynomial (due to the
presence of classical Wilson lines along flat 
geodesics) physical Hamiltonian has been obtained: it is nonlocal but without 
any kind of singularities, it has the correct Abelian limit if the structure 
constants are turned off, and it contains the explicit realization of the 
abstract Mitter-Viallet metric.

4) The Abelian and non-Abelian SU(2)
Higgs models with fermion fields\cite{lv}, where the
symplectic decoupling is a refinement of the concept of unitary gauge.
There is an ambiguity in the solutions of the Gauss law constraints, which
reflects the existence of disjoint sectors of solutions of the Euler-Lagrange
equations of Higgs models. The physical Hamiltonian and Lagrangian of  the
Higgs phase have been found; the self-energy turns out to be local and
contains a local four-fermion interaction. 

5) The standard SU(3)xSU(2)xU(1) model of elementary particles\cite{lv3}
with \hfill\break Grassmann-valued fermion fields.
The final reduced Hamiltonian contains nonlocal self-energies for the
electromagnetic and color interactions, but ``local ones" for the weak 
interactions implying the nonperturbative emergence of 4-fermions interactions.

F)
When a good description of the system in terms of Dirac observables exists,
one is going to face the problem of quantizing only the true physical
degrees of freedom, which generically are nonlinear and nonlocal functions 
or functionals of the original variables. When a quantization is possible,
there is a high probability to get a quantum theory inequivalent to that
obtained by first quantizing the original variables and then making the
reduction to the physical degrees of freedom at the quantum level (see for
instance the BRS method).

With regards to field theory, this method has the drawback that generically
the physical Hamiltonian, and therefore also the Lagrangian, is non
polynomial in the physical degrees of freedom. Power counting methods cannot
be used when looking for regularizations and renormalizations of the theory,
and the advantages of a global control of the dynamics of physical
quantities and of the possibility to check whether a model is classically
well defined
are destroyed by our present inhability to solve these problems.
The question, which puzzled both Dirac and Yukawa, reappears, whether it is
possible to define an intrinsic ultraviolet cutoff and a regularization
scheme independent from the power counting.

\section{The Separation of the Center of Mass in Special Relativity, the
Rest-Frame Instant Form of Dynamics and
Wigner-Covariant Generalized Coulomb Gauges.}

The next problem is how to covariantize these results valid in Minkowski
spacetime with Cartesian coordinates. Again the starting point 
was given by Dirac\cite{dirac} with his reformulation of classical field theory 
on spacelike hypersurfaces foliating Minkowski spacetime $M^4$ [the foliation 
is defined by an embedding $R\times \Sigma \rightarrow M^4$, $(\tau ,\vec 
\sigma ) \mapsto z^{(\mu )}(\tau ,\vec \sigma )\in \Sigma_{\tau}$, with 
$\Sigma$ an abstract 3-surface diffeomorphic to $R^3$, with $\Sigma_{\tau}$
its copy embedded in $M^4$ labelled by the value $\tau$ (the Minkowski flat 
indices are $(\mu )$; the scalar ``time" parameter $\tau$ labels
the leaves of the foliation, $\vec \sigma$ are curvilinear coordinates on
$\Sigma_{\tau}$ and $(\tau ,\vec \sigma )$ are $\Sigma_{\tau}$-adapted 
holonomic coordinates for $M^4$); this is the classical basis of
Tomonaga-Schwinger quantum field theory]. In this way one gets a parametrized 
field theory with a covariant 3+1 splitting of Minkowski spacetime and
already in a form suited to the transition to general relativity in its ADM
canonical formulation (see also Ref.\cite{kuchar}, where a 
theoretical study of this problem is done in curved spacetimes). The price is 
that one has to add as new independent configuration variables  the embedding
coordinates $z^{(\mu )}(\tau ,\vec \sigma )$ of the points 
of the spacelike hypersurface $\Sigma_{\tau}$ 
[the only ones carrying Lorentz indices] and then to define the fields on
$\Sigma_{\tau}$ so that they know  the hypersurface $\Sigma_{\tau}$ of 
$\tau$-simultaneity [for a Klein-Gordon field $\phi (x)$, this new field is
$\tilde \phi (\tau ,\vec \sigma )=\phi (z(\tau ,\vec \sigma ))$: it contains 
the nonlocal information about the embedding]. Then one rewrites the Lagrangian 
of the given isolated system in the form required by the coupling to an 
external gravitational field, makes the previous 3+1 splitting of Minkowski 
spacetime and interpretes all the fields of the system as the new fields on 
$\Sigma_{\tau}$ (they are Lorentz scalars, having only surface indices). 
Instead of considering the 4-metric as describing a 
gravitational field (and therefore as an independent field as it is done in 
metric gravity, where one adds the Hilbert action to the action for the matter 
fields), here one replaces the 4-metric with the the induced metric $g_{ AB}[z]
=z^{(\mu )}_{A}\eta_{(\mu )(\nu )}z^{(\nu )}_{B}$ on
$\Sigma_{\tau}$ [a functional of $z^{(\mu )}$;
here we use the notation $\sigma^{A}=(\tau ,\sigma^{r})$; $z^{(\mu )}_{A}=
\partial z^{(\mu )}/\partial \sigma^{A}$ are flat tetrad fields on Minkowski 
spacetime with the $z^{(\mu )}_r$'s tangent to $\Sigma_{\tau}$]
and considers the embedding coordinates $z^{(\mu )}(\tau ,\vec \sigma )$ as
independent fields [this is not possible in metric gravity, because in curved
spacetimes $z^{\mu}_{A}\not= \partial z^{\mu}/\partial \sigma^{A}$ are not
tetrad fields so that holonomic coordinates $z^{\mu}(\tau ,\vec \sigma )$
do not exist]. From this Lagrangian,
besides a Lorentz-scalar form of the constraints of the given system, 
we get four extra primary first class constraints\hfill\break
\hfill\break
${\cal H}
_{(\mu )}(\tau ,\vec \sigma )=\rho_{(\mu )}(\tau ,\vec \sigma )-l_{(\mu )}(\tau 
,\vec \sigma )T_{sys}^{\tau\tau}(\tau ,\vec \sigma )-z_{r 
(\mu )}(\tau ,\vec \sigma )T_{sys}^{\tau r}(\tau ,\vec \sigma ) 
\approx 0$\hfill\break
\hfill\break
[here $T_{sys}^{\tau\tau}(\tau ,\vec \sigma )$, $T_{sys}
^{\tau r}(\tau ,\vec \sigma )$, are the components of 
the energy-momentum tensor in the holonomic coordinate system,  
corresponding to the energy- and momentum-density of the
isolated system; one has $\lbrace {\cal H}_{(\mu )}(\tau ,\vec \sigma ),
{\cal H}_{(\nu )}(\tau ,{\vec \sigma}^{'}) \rbrace =0$]
implying the independence of the description from the choice of the 3+1 
splitting, i.e. from the choice of the foliation with spacelike hypersufaces. 
The evolution vector is given by $z^{(\mu )}_{\tau}=N_{[z](flat)}l^{(\mu )}+
N^{r}_{[z](flat)}z^{(\mu )}_{r}$, 
where $l^{(\mu )}(\tau ,\vec \sigma )$ is 
the normal to $\Sigma_{\tau}$ in $z^{(\mu )}(\tau ,\vec \sigma )$ and
$N_{[z](flat)}(\tau ,\vec \sigma )$, $N_{[z](flat)}^r(\tau ,\vec \sigma )$
are the flat lapse and shift functions defined through the metric like in 
general relativity: however, now they are not 
independent variables but functionals of $z^{(\mu )}(\tau ,\vec \sigma )$.

The Dirac Hamiltonian contains the piece 
$\int d^3\sigma \lambda^{(\mu )}(\tau ,\vec \sigma ){\cal H}_{(\mu )}
(\tau ,\vec \sigma )$ with $\lambda^{(\mu )}(\tau ,\vec \sigma )$ 
Dirac multipliers. It is possible to rewrite the integrand in the form
[${}^3g^{rs}$ is the inverse of $g_{rs}$]
\hfill\break
\hfill\break
$\lambda_{(\mu )}(\tau ,\vec 
\sigma ){\cal H}^{(\mu )}(\tau ,\vec \sigma )=[(\lambda_{(\mu )}l^{(\mu )})
(l_{(\nu )}{\cal H}^{(\nu )})-(\lambda_{(\mu )}z^{(\mu )}_{r})({}^3g
^{rs} z_{s (\nu )}{\cal H}^{(\nu )})](\tau ,\vec \sigma )$\hfill\break
${\buildrel {def} \over =}\,
N_{(flat)}(\tau ,\vec \sigma ) (l_{(\mu )}{\cal H}^{(\mu )})(\tau ,\vec \sigma )
$$-N_{(flat) r}(\tau ,\vec \sigma ) ({}^3g^{rs} z_{ s (\nu )}
{\cal H}^{(\nu )})(\tau ,\vec \sigma )$ \hfill\break
\hfill\break
with the (nonholonomic form of
the) constraints $(l_{(\mu )}{\cal H}^{(\mu )})(\tau ,\vec \sigma )\approx 0$,
$({}^3g^{rs} z_{s (\mu )} {\cal H}^{(\mu )})$\hfill\break $(\tau ,\vec 
\sigma )\approx 0$, satisfying the universal Dirac algebra of the ADM 
constraints. In this way we have defined new  flat lapse and shift functions 
\hfill\break
\hfill\break
$N_{(flat)}(\tau ,\vec \sigma )= \lambda_{(\mu )}(\tau ,\vec \sigma ) 
l^{(\mu )}(\tau ,\vec \sigma ),$\hfill\break
$N_{(flat) r}(\tau ,\vec \sigma )= \lambda_{(\mu )}(\tau ,\vec \sigma )
z^{(\mu )}_{r}(\tau ,\vec \sigma ).$\hfill\break
\hfill\break
which have the same content of the arbitrary Dirac multipliers
$\lambda_{(\mu )}(\tau ,\vec \sigma )$, namely they multiply primary
first class constraints satisfying the Dirac algebra. In 
\hfill\break Minkowski spacetime
they are quite distinct from the previous lapse and shift functions 
$N_{[z](flat)}$, $N_{[z](flat) r}$, defined starting from the metric. 
Instead in general relativity the lapse and shift functions
defined starting from the 4-metric are the coefficients (in the canonical
part $H_c$ of the Hamiltonian) of secondary first class constraints satisfying 
the Dirac algebra.

In special relativity, it is convenient to restrict ourselves to arbitrary 
spacelike hyperplanes $z^{(\mu )}(\tau ,\vec \sigma )=x^{(\mu )}_s(\tau )+
b^{(\mu )}_{r}(\tau ) \sigma^{r}$. Since they are described by 
only 10 variables, after this restriction we remain only with 10 first 
class constraints determining the 10 variables conjugate to the hyperplane 
in terms of the variables of the system:
\hfill\break
\hfill\break
${\cal H}^{(\mu )}(\tau )=p^{(\mu )}_s-p^{(\mu )}
_{(sys)}\approx 0$, ${\cal H}^{(\mu )(\nu )}(\tau )=S^{(\mu )(\nu )}
_s-S^{(\mu )(\nu )}_{(sys)}\approx 0$. \hfill\break
\hfill\break
After the restriction to spacelike hyperplanes the previous 
piece of the Dirac Hamiltonian
is reduced to ${\tilde \lambda}^{(\mu )}(\tau ){\cal H}_{(\mu )}(\tau )
-{1\over 2}{\tilde \lambda}^{(\mu )(\nu )}(\tau ){\cal H}_{(\mu 
)(\nu )}(\tau )$. Since at this stage we have $z^{(\mu )}_{r}(\tau 
,\vec \sigma )\approx b^{(\mu )}_{r}(\tau )$, so that $z^{(\mu )}
_{\tau}(\tau ,\vec \sigma )\approx N_{[z](flat)}(\tau ,\vec \sigma )l
^{(\mu )}(\tau ,\vec \sigma )+N^{r}_{[z](flat)}(\tau ,\vec \sigma )$
$b^{(\mu )}_{r}(\tau ,\vec 
\sigma )\approx {\dot x}^{(\mu )}_s(\tau )+{\dot b}^{(\mu )}_{r}(\tau )
\sigma^{r}=-{\tilde \lambda}^{(\mu )}(\tau )-{\tilde
\lambda}^{(\mu )(\nu )}(\tau )b_{r (\nu )}(\tau )\sigma^{r}$,
it is only now that we get the coincidence of the two definitions of flat
lapse and shift functions (this point was missed in the older treatments of
parametrized Minkowski theories):\hfill\break
\hfill\break
$N_{[z](flat)}(\tau ,\vec \sigma )\approx N_{(flat)}(\tau ,\vec \sigma )=
-{\tilde \lambda}
_{(\mu )}(\tau )l^{(\mu )}-l^{(\mu )}{\tilde \lambda}_{(\mu )(\nu )}(\tau )b
^{(\nu )}_{s}(\tau ) \sigma^{s},$\hfill\break
$N_{[z](flat)r}(\tau ,\vec \sigma )\approx
N_{(flat )}(\tau ,\vec \sigma )=-{\tilde \lambda}
_{(\mu )}(\tau )b^{(\mu )}_{r}(\tau )-b^{(\mu )}_{r}(\tau ){\tilde
\lambda}_{(\mu )(\nu )}(\tau ) b^{(\nu )}_{s}(\tau ) \sigma^{s}.$\hfill\break
\hfill\break
The 20 variables for the phase space description of a hyperplane are:
\hfill\break
i) $x^{(\mu )}_s(\tau ), p^{(\mu )}_s$, parametrizing the origin of 
the coordinates on the family
of spacelike hyperplanes. The four constraints ${\cal H}^{(\mu )}(\tau )
\approx 0$ say that $p_s^{(\mu )}$ is determined by the 4-momentum of the 
isolated system.\hfill\break
ii) $b^{(\mu )}_A(\tau )$ (with the $b^{(\mu )}_r(\tau )$'s being three 
orthogonal spacelike unit vectors generating the fixed $\tau$-independent
timelike unit normal $b^{(\mu )}_{\tau}=l^{(\mu )}$ to the hyperplanes) 
and $S^{(\mu )(\nu )}_s=-S^{(\nu )(\mu )}_s$ with the orthonormality 
constraints $b^{(\mu )}_A\, {}^4\eta_{(\mu )(\nu )} b^{(\nu )}_B={}^4\eta_{AB}$
[enforced by assuming the Dirac brackets 
$\{ S^{(\mu )(\nu )}_s,b^{(\rho )}_A \}={}^4\eta^{(\rho
)(\nu )} b^{(\mu )}_A-{}^4\eta^{(\rho )(\mu )} b^{(\nu )}_A$,
$\{ S^{(\mu )(\nu )}_s,S^{(\alpha )(\beta )}_s \} =C^{(\mu )(\nu )(\alpha 
)(\beta )}_{(\gamma )(\delta )} S^{(\gamma )(\delta )}_s$ 
with $C^{(\mu )(\nu 
)(\alpha )(\beta )}_{(\gamma )(\delta )}$ the structure constants of the 
Lorentz algebra]. In these variables there are hidden six independent
pairs of degrees of freedom. The six constraints ${\cal H}^{(\mu )(\nu )}
(\tau )\approx 0$ say that
$S_s^{(\mu )(\nu )}$ coincides the spin tensor of the isolated system.
Then one has that $p^{(\mu )}_s$, $J^{(\mu )(\nu )}_s=x
^{(\mu )}_sp^{(\nu )}_s-x^{(\nu )}_sp^{(\mu )}_s+S^{(\mu )(\nu )}_s$, satisfy 
the algebra of the Poincar\'e group.

Let us remark that, 
for each configuration of an isolated system there is a privileged
family of hyperplanes (the Wigner hyperplanes orthogonal to $p^{(\mu )}_s$,
existing when $ p^2_s > 0$) corresponding to the intrinsic rest-frame 
of the isolated system. If we choose these hyperplanes with suitable
gauge fixings, we remain with only  the four constraints ${\cal H}^{(\mu )}(\tau
)\approx 0$, which can be rewritten as\hfill\break
\hfill\break
$\sqrt{p^2_s} \approx [invariant\, mass\, of\, the\,
isolated\, system\, under\, investigation]= M_{sys}$; \hfill\break
${\vec p}_{sys}=[3-momentum\, of\, the\, isolated\, system\,
inside\, the\, Wigner\, hyperplane]\approx 0$.\hfill\break
\hfill\break
There is no more a restriction on $p_s^{(\mu )}$, because 
$u^{(\mu )}_s(p_s)=p^{(\mu )}_s/p^2_s$ gives the orientation of the Wigner 
hyperplanes containing the isolated system with respect to an arbitrary 
given external observer.

In this special gauge we have $b^{(\mu )}_A\equiv L^{(\mu )}{}_A(p_s,{\buildrel
\circ \over p}_s)$ (the standard Wigner boost for timelike Poincar\'e orbits),
$S_s^{(\mu )(\nu )}\equiv S_{system}^{(\mu )(\nu )}$, and the only 
remaining canonical variables are the noncovariant Newton-Wigner-like 
canonical ``external" center-of-mass
coordinate ${\tilde x}^{(\mu )}_s(\tau )$ (living on the
Wigner hyperplanes) and $p^{(\mu )}_s$.
Now 3 degrees of freedom of the isolated system [an ``internal" 
center-of-mass 3-variable ${\vec \sigma}_{sys}$ defined inside the Wigner
hyperplane and conjugate to ${\vec p}_{sys}$] become gauge variables [the
natural gauge fixing is ${\vec \sigma}_{sys}\approx 0$, so that it coincides 
with the origin $x^{(\mu )}_s(\tau )=z^{(\mu )}(\tau ,\vec \sigma =0)$ of the 
Wigner hyperplane], while the ${\tilde x}^{(\mu )}$ 
is playing the role of a kinematical external
center of mass for the isolated system and may be interpreted as a decoupled 
observer with his parametrized clock (point particle clock).
All the fields living on the Wigner hyperplane are now either Lorentz scalar 
or with their 3-indices transformaing under Wigner rotations (induced by Lorentz
transformations in Minkowski spacetime) as any Wigner spin 1 index.

One obtains 
in this way a new kind of instant form of the dynamics (see Ref.\cite{dira2}), 
the  ``Wigner-covariant 1-time rest-frame instant form"\cite{lus1} with a 
universal breaking of Lorentz covariance. 
It is the special relativistic generalization of
the nonrelativistic separation of the center of mass from the relative motion
[$H={{ {\vec P}^2}\over {2M}}+H_{rel}$]. The role of the center of mass is 
taken by the Wigner hyperplane, identified by the point ${\tilde x}^{(\mu )}
(\tau )$ and by its normal $p^{(\mu )}_s$. The
invariant mass $M_{sys}$ of the system replaces the nonrelativistic  Hamiltonian
$H_{rel}$ for the relative degrees of freedom, after the addition of the
gauge-fixing $T_s-\tau \approx 0$ [identifying the time parameter $\tau$,
labelling the leaves of the foliation,  with 
the Lorentz scalar time of the center of mass in the rest frame,
$T_s=p_s\cdot {\tilde x}_s/M_{sys}$; $M_{sys}$  generates the
evolution in this time]. 

The determination of ${\vec \sigma}_{sys}$ may be done with the group 
theoretical methods of Ref.\cite{pauri}: given a realization on the phase space
of a given system of the ten Poincar\'e generators one can build three 
3-position variables only in terms of them, which in our case of a system
on the Wigner hyperplane with ${\vec p}_{sys}\approx 0$ are: i) a canonical 
center of mass (the ``internal" center of mass ${\vec \sigma}_{sys}$); ii)
a noncanonical M\o ller center of energy ${\vec \sigma}^{(E)}_{sys}$; iii)
a noncanonical Fokker-Pryce center of inertia ${\vec \sigma}^{(FP)}_{sys}$. Due 
to ${\vec p}_{sys}\approx 0$, we have ${\vec \sigma}_{sys} \approx
{\vec \sigma}^{(E)}_{sys} \approx {\vec \sigma}^{(FP)}_{sys}$. By adding the
gauge fixings ${\vec \sigma}_{sys}\approx 0$ one can show that the origin
$x_s^{(\mu )}(\tau )$ becomes  simultaneously the Dixon center of mass of
an extended object and both the Pirani and Tulczyjew centroids (see Ref.
\cite{mate} for the application of these methods to find the center of mass
of a configuration of the Klein-Gordon field after the preliminary work of
Ref.\cite{lon}). With similar methods one can construct three ``external"
collective positions (all located on the Wigner hyperplane): i) the ``external"
canonical noncovariant center of mass ${\tilde x}_s^{(\mu )}$; ii) the
``external" noncanonical and noncovariant M\o ller center of energy 
$R^{(\mu )}_s$; iii) the ``external" covariant noncanonical Fokker-Pryce center 
of inertia $Y^{(\mu )}_s$ (when
there are the gauge fixings ${\vec \sigma}_{sys}\approx 0$ it also coincides
with the origin $x^{(\mu )}_s$). It turns out that the Wigner hyperplane is
the natural setting for the study of the Dixon multipoles of extended 
relativistic systems\cite{dixon} and for defining the canonical relative
variables with respect to the center of mass. After having put control on the
relativistic definitions of center of mass of an extended system, the lacking
kinematics of relativistic rotations in now under investigation.
The Wigner hyperplane with its 
natural Euclidean metric structure offers a natural solution to the problem of
boost for lattice gauge theories and realizes explicitly the machian aspect of
dynamics that only relative motions are relevant.

The isolated systems till now analyzed to get their rest-frame 
Wigner-covariant generalized
Coulomb gauges [i.e. the subset of global Shanmugadhasan canonical bases, 
which, for each Poincar\'e stratum, are also adapted to the geometry of the
corresponding Poincar\'e orbits with their little groups; these special bases
can be named Poincar\'e-Shanmugadhasan bases for the given Poincar\'e stratum
of the presymplectic constraint manifold (every stratum requires an independent
canonical reduction); till now only the main stratum with
$p^2$ timelike and $W^2\not= 0$ has been investigated] are:

a) The system of N scalar particles with Grassmann electric charges
plus the electromagnetic field \cite{lus1}. The starting configuration 
variables are a 3-vector ${\vec \eta}_i(\tau )$ for each particle [$x^{(\mu )}
_i(\tau )=z^{(\mu )}(\tau ,{\vec \eta}
_i(\tau ))$] and the electromagnetic gauge potentials 
$A_{A}(\tau ,\vec \sigma )={{\partial z^{(\mu )}(\tau ,\vec \sigma )}
\over {\partial \sigma^{A}}} A_{(\mu )}(z(\tau ,\vec \sigma ))$, 
which know  the embedding of
$\Sigma_{\tau}$ into $M^4$. One has to choose the sign of the energy of each
particle, because there are not mass-shell constraints (like $p_i^2-m^2_i\approx
0$) among the constraints of this formulation, due to the fact that one has only
three degrees of freedom for particle, determining the intersection of a 
timelike trajectory and of the spacelike hypersurface $\Sigma_{\tau}$. For
each choice of the sign of the energy of the N particles, one describes only one
of the $2^N$ branches of the mass spectrum of the manifestly covariant approach 
based on the coordinates $x^{(\mu )}_i(\tau )$, $p^{(\mu )}_i(\tau )$, 
i=1,..,N, and on
the constraints $p^2_i-m^2_i\approx 0$ (in the free case). In this way, one 
gets a description of relativistic particles with a given sign of the energy
with consistent couplings to fields and valid independently from the quantum
effect of pair production [in the manifestly covariant approach, containing
all possible branches of the particle mass spectrum, the classical counterpart 
of pair production is the intersection of different branches deformed by the
presence of interactions]. The final Dirac's observables are: i) the transverse 
radiation field variables ${\vec A}_{\perp}$, ${\vec E}_{\perp}$; 
ii) the particle canonical variables ${\vec \eta}_i(\tau )$, ${{\vec \kappa}}
_i(\tau )$, dressed with a Coulomb cloud. The physical Hamiltonian contains the 
mutual instantaneous Coulomb 
potentials extracted from field theory and there is a regularization of the
Coulomb self-energies due to the Grassmann character of the electric charges
$Q_i$ [$Q^2_i=0$]. In Ref.\cite{lus2} there is the study of the 
Lienard-Wiechert potentials and of Abraham-Lorentz-Dirac equations in this
rest-frame Coulomb gauge and also scalar electrodynamics is reformulated in it.
Also the rest-frame 1-time relativistic statistical mechanics has been developed
\cite{lus1}.

b) The system of N scalar particles with Grassmann-valued color charges plus 
the color SU(3) Yang-Mills field\cite{lus3}: 
it gives the pseudoclassical description of the
relativistic scalar-quark model, deduced from the classical QCD Lagrangian and 
with the color field present. The physical invariant mass of the system is
given in terms of the Dirac observables. From the reduced Hamilton equations  
the second order equations of motion both for the reduced transverse color 
field and the particles are extracted. Then, one studies  the N=2 
(meson) case. A special form of the requirement of having only color singlets, 
suited for a field-independent quark model, produces a ``pseudoclassical 
asymptotic freedom" and a regularization of the quark self-energy. With these
results one can covariantize the bosonic part of the standard model given in
Ref.\cite{lv3}.
 
c) The system of N spinning particles of definite energy [$({1\over 2},0)$ or
$(0,{1\over 2})$ representation of SL(2,C)] with Grassmann electric charges 
plus the electromagnetic field\cite{biga} and that of a Grassmann-valued
Dirac field plus the electromagnetic field (the pseudoclassical basis of QED) 
\cite{bigaz}. In both cases there are geometrical complications connected with 
the spacetime description of the path of electric currents and not only of their
spin structure, suggesting a reinterpretation of the supersymmetric scalar 
multiplet as a spin fibration with the Dirac field in the fiber and the
Klein-Gordon field in the base; a new canonical decomposition of the 
Klein-Gordon field into center-of-mass and relative variables \cite{lon,mate} 
will be helpful to clarify these problems. After their solution and after having
obtained the description of Grassmann-valued chiral fields [this will require
the transcription of the front form of the dynamics in the instant one for the
Poincar\'e strata with $P^2=0$] the rest-frame form of the full standard 
$SU(3)\times SU(2)\times U(1)$ model can be achieved.

The rest-frame description of the relativistic perfect gas is now under 
investigation.

All these new pieces of information  will allow, after quantization of this new
consistent relativistic mechanics without the classical problems connected
with pair production, to find the  asymptotic states of the covariant
Tomonaga-Schwinger formulation of quantum field theory on spacelike
hypersurfaces (to be obtained by quantizing the fields on $\Sigma_{\tau}$):
these states are needed for the theory of quantum bound states
[since Fock states do not constitute a Cauchy problem for the field equations,
because an in (or out) particle can be in the absolute future of another one due
to the tensor product nature of these asymptotic states, bound state equations
like the Bethe-Salpeter one have spurious solutions which are excitations in
relative energies, the variables conjugate to relative times]. Moreover, it 
will be possible to include bound states among the asymptotic states.

As said in Ref.\cite{lus2,lus3}, the quantization of these rest-frame
models has to overcome two problems. On the particle
side, the complication is the quantization of the square roots associated
with the relativistic kinetic energy terms: in the free case this has been done
in Ref.\cite{lam} [see Refs.\cite{sqroot} for the complications induced by the
Coulomb potential]. On the field side (all physical
Hamiltonian are nonlocal and, with the exception of the Abelian case,
nonpolynomial, but quadratic in the momenta), the obstacle
is the absence (notwithstanding there is no  no-go theorem) of a complete
regularization and renormalization procedure of electrodynamics (to start with) 
in the Coulomb gauge: see Ref.\cite{cou} (and its bibliography)
for the existing results for QED.

However, as shown in Refs.\cite{lus1,lusa}, the rest-frame instant 
form of dynamics automatically gives a physical ultraviolet cutoff in the 
spirit of Dirac and Yukawa: it is the M$\o$ller radius\cite{mol} 
$\rho =\sqrt{-W^2}/p^2=|\vec S|/\sqrt{p^2}$ ($W^2=-p^2{\vec 
S}^2$ is the Pauli-Lubanski Casimir when $p^2 > 0$), namely the classical 
intrinsic radius of the worldtube, around the covariant noncanonical 
Fokker-Pryce center of inertia $Y^{(\mu )}$, 
inside which the noncovariance of the canonical center of mass ${\tilde
x}^{\mu}$ is concentrated. At the quantum level $\rho$ becomes the Compton 
wavelength of the isolated system multiplied its spin eigenvalue $\sqrt{s(s+1)}$
, $\rho \mapsto \hat \rho = \sqrt{s(s+1)} \hbar /M=\sqrt{s(s+1)} \lambda_M$ 
with $M=\sqrt{p^2}$ the invariant mass and $\lambda_M=\hbar /M$ its Compton
wavelength. Therefore, the criticism to classical relativistic physics, based
on quantum pair production, concerns the testing of distances where, due to the
Lorentz signature of spacetime, one has intrinsic classical covariance problems:
it is impossible to localize the canonical center of mass ${\tilde x}^{\mu}$
adapted to the first class constraints of the system
(also named Pryce center of mass and having the same covariance of the 
Newton-Wigner position operator) in a frame independent way.

Let us remember \cite{lus1}
that $\rho$ is also a remnant in flat Minkowski spacetime of 
the energy conditions of general relativity: since the M$\o$ller noncanonical, 
noncovariant center of energy $R^{(\mu )}$has its noncovariance localized
inside the same worldtube with radius $\rho$ (it was discovered in this way)
\cite{mol}, it turns out that for an extended relativistic system with the
material radius smaller of its intrinsic radius $\rho$ one has: i) its 
peripheral rotation velocity can exceed the velocity of light; ii) its 
classical energy density cannot be positive definite everywhere in every frame. 

Now, the real relevant point is that this ultraviolet cutoff determined by
$\rho$ exists also in Einstein's
general relativity (which is not power counting renormalizable) in the case of
asymptotically flat spacetimes, taking into account the Poincar\'e Casimirs of
its asymptotic ADM Poincar\'e charges (when supertranslations are eliminated 
with suitable boundary conditions). The generalization of the 
worldtube of radius $\rho$ to asymptotically flat general relativity with 
matter, could also be connected with the unproved cosmic censorship hypothesis.

Moreover, the extended Heisenberg relations  of string theory\cite{ven}, i.e.
$\triangle x ={{\hbar}\over {\triangle p}}+{{\triangle p}\over {T_{cs}}}=
{{\hbar}\over {\triangle p}}+{{\hbar \triangle p}\over {L^2_{cs}}}$ implying the
lower bound $\triangle x > L_{cs}=\sqrt{\hbar /T_{cs}}$ due to the $y+1/y$
structure,
have a counterpart in the quantization of the M$\o$ller radius\cite{lus1}:
if we ask that, also at the quantum level, one cannot test the inside of the 
worldtube, we must ask $\triangle x > \hat \rho$ which is the lower bound
implied by the modified uncertainty relation $\triangle x ={{\hbar}\over 
{\triangle p}}+{{\hbar \triangle p}\over {{\hat \rho}^2}}$. This could imply 
that the center-of-mass canonical noncovariant  3-coordinate 
$\vec z=\sqrt{P^2}({\vec {\tilde x}}-{{\vec P}\over {P^o}}{\tilde x}^o)$ 
\cite{lus1} cannot become a
self-adjoint operator. See Hegerfeldt's theorems (quoted in 
Refs.\cite{lusa,lus1}) and his interpretation 
pointing at the impossibility of a good localization of relativistic particles
(experimentally one determines only a worldtube in spacetime emerging from the 
interaction region). Since the eigenfunctions of the canonical center-of-mass
operator are playing the role of the wave function of the universe, one could 
also say that the center-of-mass variable has not to be quantized, because it
lies on the classical macroscopic side of Copenhagen's interpretation and,
moreover, because, in the spirit of Mach's principle that only relative 
motions can be observed, no one can observe it (it is only used to define a
decoupled ``point particle clock"). On the other hand, if one 
rejects the canonical noncovariant center of mass in favor of the covariant
noncanonical Fokker-Pryce center of inertia $Y^{\mu}$, $\{ Y^{\mu},Y^{\nu} \}
\not= 0$, one could invoke the philosophy of quantum groups to quantize 
$Y^{\mu}$ to get some kind of quantum plane for the center-of-mass 
description. Let us remark that the quantization of the square root Hamiltonian
done in Ref.\cite{lam} is consistent with this problematic.

In conclusion, the best set of canonical coordinates adapted to the constraints
and to the geometry of Poincar\'e orbits in Minkowski spacetime
and naturally predisposed to the
coupling to canonical tetrad gravity is emerging for the electromagnetic, weak
and strong interactions with matter described either by fermion fields or by
relativistic particles with a definite sign of the energy.

\section{Tetrad Gravity, Physical Hamiltonian Degrees of Freedom of the
Gravitational Field and the Deparametrization of General Relativity.}

Tetrad gravity is the formulation of general relativity natural for the
coupling to the fermion fields of the standard model. However, we need a
formulation of it, which allows to solve its constraints for doing the
canonical reduction and to solve
the deparametrization problem of general relativity (how to recover
the rest-frame instant form when the Newton constant is put equal to zero,
G=0). Since neither a complete reduction
of gravity with an identification of the physical canonical degrees of
freedom of the gravitational field nor a detailed study of its Hamiltonian
group of gauge transformations (whose infinitesimal generators are the first
class constraints) has ever been pushed till the end in an explicit way,
a new formulation of tetrad gravity 
\cite{russo1,russo2,russo3,russo4} was developed.

To implement this program
we shall restrict ourselves to the simplest class of spacetimes 
[time-oriented pseudo-Riemannian or Lorentzian 4-manifold $(M^4,{}^4g)$ with 
signature $\epsilon \, (+---)$ ($\epsilon =\pm 1$ according to either particle
physics or general relativity convention) and with a choice of time 
orientation], assumed to be:

i) Globally hyperbolic 4-manifolds, i.e. topologically they are $M^4=R\times 
\Sigma$, so to have a well posed Cauchy problem [with $\Sigma$ the abstract
model of Cauchy surface] at least till when no singularity develops in $M^4$
[see the singularity theorems]. Therefore, these spacetimes admit regular 
foliations with orientable, complete, non-intersecting spacelike 3-manifolds
$\Sigma_{\tau}$ [$\tau :M^4 
\rightarrow R$, $z^{\mu} \mapsto \tau (z^{\mu})$, is a global timelike
future-oriented function labelling the leaves (surfaces of simultaneity)]. In 
this way, one obtains 3+1 splittings of $M^4$ and the possibility of a 
Hamiltonian formulation.

ii) Asymptotically flat at spatial infinity, so to have the possibility to 
define asymptotic Poincar\'e charges \cite{adm,reg,ash}: they 
allow the definition of a M$\o$ller radius also in general relativity and are a
bridge towards a future soldering with the theory of elementary particles in
Minkowski spacetime defined as irreducible representation of its kinematical,
globally implemented Poincar\'e group according to Wigner. This excludes
Einstein-Wheeler closed universes without boundaries (no asymptotic
Poincar\'e charges), which were introduced to eliminate boundary conditions at
spatial infinity to make the theory as machian as possible. 

iii) Admitting a spinor (or spin) structure\cite{wald} for the coupling
to fermion fields. Since we 
consider noncompact space- and time-orientable spacetimes, spinors can be
defined if and only if they are ``parallelizable" \cite{ger}, like in our case.
This implies that the orthonormal frame principal SO(3)-bundle over 
$\Sigma_{\tau}$ (whose connections are the spin connections determined by the 
cotriads) is trivial.

iv) The noncompact parallelizable simultaneity 3-manifolds (the Cauchy surfaces)
$\Sigma_{\tau}$ are assumed to be topologically trivial, geodesically complete
and, finally, diffeomorphic to
$R^3$. These 3-manifolds have the same manifold structure as Euclidean spaces:
a) the geodesic exponential map $Exp_p:T_p\Sigma_{\tau}\rightarrow \Sigma
_{\tau}$ is a diffeomorphism ; b) the sectional curvature is 
less or equal  zero everywhere; c) they have no ``conjugate locus" [i.e.
there are no pairs of conjugate Jacobi points (intersection points of distinct
geodesics through them) on any geodesic] and no ``cut locus" [i.e. no closed
geodesics through any point].

v) Like in Yang-Mills case \cite{lusa}, the 3-spin-connection on the orthogonal
frame SO(3)-bundle (and therefore cotriads) will have to be 
restricted to suited weighted Sobolev spaces to avoid Gribov ambiguities
\cite{lusa,moncr}. In
turn, this implies the absence of isometries of the noncompact Riemannian
3-manifold $(\Sigma_{\tau},{}^3g)$ [see for instance the review paper in Ref.
\cite{cho}]. 

Diffeomorphisms on $\Sigma_{\tau}$ ($Diff\, \Sigma_{\tau}$) are interpreted 
in the passive way, following Ref.\cite{be}, in accord with the Hamiltonian
point of view that infinitesimal diffeomorphisms are generated by taking the
Poisson bracket with the 1st class supermomentum constraints [passive
diffeomorphisms are also named `pseudodiffeomorphisms'].

The new formulation of tetrad gravity [see Refs.
\cite{weyl} for the existing
versions of the theory] utilizes the ADM action of metric gravity with the
4-metric expressed in terms of arbitrary cotetrads. Let us remark that both
in the ADM metric and tetrad formulation one has to introduce the extra
ingredient of the 3+1 splittings of $M^4$ with foliations whose leaves $\Sigma
_{\tau}$ are spacelike 3-hypersurfaces. However, their points $z^{\mu}(\tau
,\vec \sigma )$ [$(\tau ,\vec \sigma )$ are $\Sigma_{\tau}$-adapted holonomic
coordinates of $M^4$] are not configurational variables of these theories in
contrast to what happens in Minkowski parametrized theories 
as already said [${{\partial 
z^{\mu}}\over {\partial \sigma^A}}$ are not tetrads when $M^4$ is not
Minkowski spacetime with Cartesian coordinates, because ${}^4g^{AB}{{\partial 
z^{\mu}}\over {\partial \sigma^A}} {{\partial z^{\nu}}\over {\partial 
\sigma^B}}={}^4g^{\mu\nu}\not= {}^4\eta^{(\mu )(\nu )}$].

By using $\Sigma_{\tau}$-adapted holonomic coordinates for $M^4$, one has 
found a new \hfill\break
parametrization of arbitrary tetrads and cotetrads on $M^4$ in 
terms of cotriads on $\Sigma_{\tau}$ [${}^3e_{(a)r}(\tau ,\vec \sigma )$], of
lapse [$N(\tau ,\vec \sigma )$] and shift [$N_{(a)}(\tau ,\vec \sigma )=
\{{}^3e_{(a)r} N^r\} (\tau ,\vec \sigma )$] functions and of 3 parameters
[$\varphi_{(a)}(\tau ,\vec \sigma )$] parametrizing point-dependent Wigner 
boosts for timelike Poincar\'e orbits. Putting these variables in the ADM 
action for metric gravity \cite{adm} (with the 3-metric on
$\Sigma_{\tau}$ expressed in terms of cotriads: ${}^3g_{rs}={}^3e_{(a)r}\,
{}^3e_{(a)s}$ with positive signature), one gets a new action
depending only on lapse, shifts and cotriads, but not on the boost parameters
(therefore, there is no need to use Schwinger's time gauge). There are 10
primary and 4 secondary first class constraints and a weakly vanishing canonical
Hamiltonian containing the secondary constraints like in ADM metric gravity
\cite{adm}. Besides the 3 constraints associated with the vanishing
Lorentz boost momenta (Abelianization of boosts), there are
4 constraints saying that the momenta associated with lapse and shifts vanish,
3 constraints describing rotations, 3 constraints generating 
space-diffeomorphisms on the cotriads induced by those ($Diff\, \Sigma_{\tau}$)
on $\Sigma_{\tau}$ (a linear combination of supermomentum constraints and
of the rotation ones;a different combination of these constraints generates
SO(3) Gauss law constraints for the momenta ${}^3{\tilde \pi}^r_{(a)}$
conjugated to cotriads with the covariant derivative built with the spin
connection) and one superhamiltonian constraint. The six constraints connected 
with Lorentz boosts and rotations replace the constraints satisfying the 
Lorentz algebra in the older formulations. The boost parameters $\varphi
_{(a)}(\tau ,\vec \sigma )$ and the three angles $\alpha_{(a)}(\tau ,\vec 
\sigma )$ hidden in the cotriads are the extra variables of tetrad gravity with 
respect to metric gravity: they allow a Hamiltonian description of the 
congruences of timelike accelerated observers used in the formulation of 
gravitomagnetism\cite{bini,ciuf}.

It turns out that 
with the technology developed for Yang-Mills theory, one can Abelianize the 3 
rotation constraints and then also the space-diffeomorphism constraints so
that we can arrive at a total of 13 Abelianized first class constraints. 
In the Abelianization of the rotation constraints one needs the Green function
of the 3-dimensional covariant derivative containing the spin connection,
well defined only if there is no Gribov ambiguity in the SO(3)-frame bundle 
and no isometry of the Riemannian 3-manifold
$(\Sigma_{\tau},{}^3g)$. The Green function is similar
to the Yang-Mills one for a principal SO(3)-bundle \cite{lusa}, but, instead
of the Dirac distribution for the Green function of the flat
divergence, it contains the Synge-DeWitt  bitensor \cite{dew} defining
the tangent in one endpoint of the geodesic arc connecting two points (which 
reduces to the 
Dirac distribution only locally in normal coordinates). Moreover, the definition
of the Green function now requires the geodesic exponential map.

In the resulting quasi-Shanmugadhasan canonical basis, 
the original cotriad can be expressed in closed
form in terms of 3 rotation angles, 3 diffeomorphism-parameters and a
reduced cotriad depending only on 3 independent variables
(they are Dirac's observables with respect to 13 of the 14 first class 
constraints) and with their conjugate momenta, still subject to the reduced 
form of the superhamiltonian constrain: this is the phase space over the 
superspace of 3-geometries\cite{witt}.

Till now no coordinate condition\cite{cji} has been imposed. It turns out that
these conditions are hidden in the choice of how to parametrize the reduced
cotriads in terms of three independent functions. The
simplest parametrization (the only one studied till now) corresponds to choose a
system of global 3-orthogonal coordinates on $\Sigma_{\tau}$, in which the 
3-metric is diagonal. With a further canonical transformation on the reduced 
cotriads and conjugate momenta, one arrives at a canonical basis containing the 
conformal factor $\phi (\tau ,\vec \sigma )=e^{q(\tau ,\vec \sigma )/2}$ of the 
3-geometry and its conjugate momentum $\rho (\tau ,\vec \sigma )$ plus two 
other pairs of conjugate canonical variables $r_{\bar a}(\tau ,\vec \sigma )$,
$\pi_{\bar a}(\tau ,\vec \sigma )$, $\bar a = 1,2$. 
The reduced superhamiltonian constraint, expressed in terms of these variables, 
turns out to be an integro-differential 
equation for the conformal factor (reduced Lichnerowicz equation) whose 
conjugate momentum is, therefore, the last gauge variable. If we replace the
gauge fixing of the Lichnerowicz\cite{conf} and York\cite{york,yoyo,ciuf}
approach [namely the vanishing of the trace of the extrinsic curvature of
$\Sigma_{\tau}$, ${}^3K(\tau ,\vec \sigma )\approx 0$, also named the internal
extrinsic York time\cite{qadir}] with the natural one $\rho (\tau ,\vec \sigma )
\approx 0$ and we go to Dirac brackets, we find that $r_{\bar a}(\tau ,\vec 
\sigma )$, $\pi_{\bar a}(\tau ,\vec \sigma )$ are the canonical basis for the 
physical degrees of freedom or Dirac's observables of the gravitational field
in the 3-orthogonal gauges. Let us remark that the functional form of the
non-tensorial objects $r_{\bar a}$, $\pi_{\bar a}$, depends on the chosen 
coordinate condition.

The next step is to find the physical Hamiltonian for them and to solve the
deparametrization problem.
If we wish to arrive at the soldering
of tetrad gravity with matter and parametrized Minkowski formulation for the 
same matter, we must require that the lapse and shift functions of tetrad 
gravity [which must grow linearly in $\vec \sigma$, in suitable asymptotic
Minkowski coordinates, according to the existing literature on asymptotic
Poincar\'e charges at spatial infinity \cite{reg}] 
must agree asymptotically with the flat lapse and shift functions, which,
however, are unambigously defined only on Minkowski spacelike hyperplanes
as we have seen.

In metric ADM gravity the canonical Hamiltonian is $H_{(c)ADM}=\int d^3\sigma
[N {\tilde {\cal H}}+N_r {\tilde {\cal H}}^r](\tau ,\vec \sigma )\approx 0$,
where ${\tilde {\cal H}}(\tau ,\vec \sigma )\approx 0$ and ${\tilde {\cal H}}
^r(\tau ,\vec \sigma )\approx 0$ are the superhamiltonian and supermomentum 
constraints. It is differentiable and finite only for suitable $N(\tau ,\vec 
\sigma )=n(\tau ,\vec \sigma )\, {\rightarrow}_{|\vec \sigma |\rightarrow 
\infty}\, 0$, $N_r(\tau ,\vec \sigma )=n_r(\tau ,\vec \sigma )\, {\rightarrow}
_{|\vec \sigma |\rightarrow \infty}\, 0$ defined  by Beig and
\'O'Murchadha\cite{reg} in
suitable asymptotic coordinate systems. For more general lapse and shift
functions one must add a surface term \cite{witt} to $H_{(c)ADM}$, which 
contains the ``strong" Poincar\'e charges \cite{adm} $P^A_{ADM}$,
$J^{AB}_{ADM}$ [they are conserved and 
gauge invariant surface integrals]. To have well defined asymptotic Poincar\'e
charges at spatial infinity\cite{adm,reg} one needs: i) the selection
of a class of coordinates systems for $\Sigma_{\tau}$ asymptotic to flat 
coordinates; ii) the choice of a class of Hamiltonian boundary conditions 
for the fields in
these coordinate systems [all the fields must belong to some functional space 
of the type of the weighted Sobolev spaces]; iii) a definition of the 
Hamiltonian group ${\cal G}$ of gauge transformations (and in particular of
proper gauge transformations) with a well defined 
limit at  spatial infinity so to respect i) and ii). The scheme is the same
needed to define the non-Abelian charges in Yang-Mills theory\cite{lusa}. The
delicate point is to be able to exclude supertranslations\cite{wald}, because
the presence of these extra asymptotic charges leads to the replacement of the
asymptotic Poincar\'e group with the infinite-dimensional spi group\cite{ash}
of asymptotic symmetries,
which does not allow the definition of the Poincar\'e spin due to the absence of
the Pauli-Lubanski Casimir. This can be done with suitable boundary conditions
(in particular all the fields and gauge transformations must have
direction independent limits at spatial infinity) respecting the ``parity 
conditions" of Beig and \'O'Murchadha\cite{reg}.

Let us then remark that
in Ref.\cite{drc} and in the book in Ref.\cite{dirac} (see also Ref.\cite{reg}),
Dirac introduced asymptotic Minkowski rectangular coordinates\hfill\break
\hfill\break
$z^{(\mu )}_{(\infty )}(\tau ,\vec \sigma )=x^{(\mu )}_{(\infty )}(\tau )+
b^{(\mu )}_{(\infty )\, r}(\tau ) \sigma^{r}$ \hfill\break
\hfill\break
in $M^4$ at spatial infinity $S_{\infty}=\cup_{\tau} S^2_{\tau ,\infty}$ 
For each value of $\tau$, the coordinates $x^{(\mu )}
_{(\infty )}(\tau )$ labels a point, near spatial infinity chosen as origin of
$\Sigma_{\tau}$. On it there is a flat tetrad $b^{(\mu )}_{(\infty )\, A}
(\tau )= \{ \, l^{(\mu )}_{(\infty )}=b^{(\mu )}_{(\infty )\, \tau}=\epsilon
^{(\mu )}{}_{(\alpha )(\beta )(\gamma )} b^{(\alpha )}_{(\infty )\, 1}
(\tau )b^{(\beta )}_{(\infty )\, 2}(\tau )b^{(\gamma )}_{(\infty )\, 
3}(\tau );$\hfill\break
$\, b^{(\mu )}_{(\infty )\, r}(\tau )\, \}$, with
$l^{(\mu )}_{(\infty )}$ $\tau$-independent, satisfying $b^{(\mu )}_{(\infty )\,
A}\, {}^4\eta_{(\mu )(\nu )}\, b^{(\nu )}_{(\infty )\, B}={}^4\eta_{AB}$ for
every $\tau$ [at this level we do not assume that $l^{(\mu )}_{(\infty )}$ is
tangent to $S_{\infty}$, as the normal $l^{\mu }$ to $\Sigma_{\tau}$]. 
There will be transformation coefficients $b^{\mu}_A(\tau ,\vec 
\sigma )$ from the holonomic adapted coordinates $\sigma^A=(\tau ,\sigma
^{r})$ to coordinates $x^{\mu}=z^{\mu}(\sigma^A)$ in an atlas of $M^4$,
such that in a chart at spatial infinity one has $z^{\mu}(\tau ,\vec \sigma )
=\delta^{\mu}_{(\mu )} z^{(\mu )}(\tau ,\vec \sigma )$ and $b^{\mu}_A(\tau 
,\vec \sigma )= \delta^{\mu}_{(\mu )} b^{(\mu )}_{(\infty )A}(\tau )$
[for $r\, \rightarrow \, \infty$ one has ${}^4g_{\mu\nu}\, \rightarrow \,
\delta^{(\mu )}_{\mu}\delta^{(\nu )}_{\nu}{}^4\eta_{(\mu )(\nu )}$ and
${}^4g_{AB}=b^{\mu}_A\, {}^4g_{\mu\nu} b^{\nu}_B\, \rightarrow \,
b^{(\mu )}_{(\infty )A}\, {}^4\eta_{(\mu )(\nu )} b^{(\nu )}_{(\infty )B}=
{}^4\eta_{AB}$ ].

Dirac\cite{drc} and, then, Regge and
Teitelboim\cite{reg} proposed that the asymptotic Minkowski rectangular
coordinates $z^{(\mu )}_{(\infty )}(\tau ,\vec \sigma )=x^{(\mu )}_{(\infty )}
(\tau )+b^{(\mu )}_{(\infty ) r}(\tau )\sigma^{r}$ should define 
10 new independent degrees of freedom at the spatial boundary $S_{\infty}$, 
as it happens for Minkowski parametrized theories\cite{lus1} when restricted 
to spacelike hyperplanes [defined by $z^{(\mu )}(\tau ,\vec \sigma )\approx 
x^{(\mu )}_s(\tau )+b^{(\mu )}_{r}(\tau )\sigma^{r}$]; then, 
10 conjugate momenta should exist.
These 20 extra variables of the Dirac proposal
can be put in the form: $x^{(\mu )}_{(\infty )}(\tau )$,
$p^{(\mu )}_{(\infty )}$, $b^{(\mu )}_{(\infty ) A}(\tau )$ [with $b^{(\mu )}
_{(\infty ) \tau }=l^{(\mu )}_{(\infty )}$ $\tau$-independent], $S^{(\mu )(\nu 
)}_{(\infty )}$, with  Dirac brackets implying the orthonormality
constraints $b^{(\mu )}_{(\infty ) A}\, {}^4\eta_{(\mu )(\nu )} b^{(\nu )}
_{(\infty ) B}={}^4\eta_{AB}$ [so that $p^{(\mu )}_{(\infty )}$ and 
$J^{(\mu )(\nu )}_{(\infty )}=x^{(\mu )}_{(\infty )}p^{(\nu )}_{(\infty )}-
x^{(\nu )}_{(\infty )}p^{(\mu )}_{(\infty )}+S^{(\mu )(\nu )}_{(\infty )}$
satisfy a Poincar\'e algebra]. In analogy with Minkowski parametrized
theories restricted to spacelike hyperplanes, one expects to have 10 extra
first class constraints of the type \hfill\break
\hfill\break
$p^{(\mu )}_{(\infty )}-P^{(\mu )}
_{ADM}\approx 0$, $S^{(\mu )(\nu )}_{(\infty )}-S^{(\mu )(\nu )}_{ADM}
\approx 0$ \hfill\break
\hfill\break
with $P^{(\mu )}_{ADM}$, $S^{(\mu )(\nu )}_{ADM}$ related to the
ADM Poincar\'e charges $P^A_{ADM}$, $J^{AB}_{ADM}$.
The origin $x^{(\mu )}_{(\infty )}$ is going to play
the role of a decoupled observer with his parametrized clock.

Let us remark that if we
replace $p^{(\mu )}_{(\infty )}$ and $S^{(\mu )(\nu )}_{(\infty )}$, whose
Poisson algebra is the direct sum of an Abelian algebra of translations and of 
a Lorentz algebra, with the new variables (with holonomic indices with
respect to $\Sigma_{\tau}$) $p^A_{(\infty )}=b^A_{(\infty )(\mu )}p^{(\mu )}
_{(\infty )}$, $J^{AB}_{(\infty )}=b^A_{(\infty )(\mu )}b^B_{(\infty )(\nu )}
S^{(\mu )(\nu )}_{(\infty )}$ [$\not= b^A_{(\infty )(\mu )}b^B_{(\infty )(\nu )}
J^{(\mu )(\nu )}_{(\infty )}$], the Poisson brackets for $p^{(\mu )}
_{(\infty )}$, $b^{(\mu )}_{(\infty ) A}$, $S^{(\mu )(\nu )}_{(\infty )}$ 
imply that $p^A_{(\infty )}$, $J^{AB}_{(\infty )}$ satisfy a Poincar\'e
algebra. This  implies that the Poincar\'e generators ${P}^A_{ADM}$, ${J}^{AB}
_{ADM}$ define in the asymptotic Dirac rectangular coordinates a momentum 
${P}^{(\mu )}_{ADM}$ and only an  ADM spin tensor ${S}^{(\mu )(\nu )}_{ADM}$
[to define an angular momentum tensor $J^{(\mu )(\nu )}_{ADM}$ one should
find a ``center of mass of the gravitational field" $X^{(\mu )}_{ADM} [{}^3g,
{}^3{\tilde \Pi}]$ (see Ref.\cite{lon} for the Klein-Gordon case) conjugate to
$P^{(\mu )}_{ADM}$, so that $J^{(\mu )(\nu )}_{ADM}=X^{(\mu )}_{ADM}P^{(\nu )}
_{ADM}-X^{(\nu )}_{ADM}P^{(\mu )}_{ADM}+S^{(\mu )(\nu )}_{ADM}$].

The following splitting of the lapse and shift functions and the following set
of boundary conditions fulfill all the previous requirements [soldering with
the lapse and shift functions on Minkowski hyperplanes; absence of 
supertranslations [strictly speaking one gets $P^r_{ADM}=0$ due to the parity
conditions; $r=|\vec \sigma |$]\hfill\break
\hfill\break
${}^3g_{rs}(\tau ,\vec \sigma )\, {\rightarrow}_{r\, 
\rightarrow \infty}\, (1+{M\over r})
\delta_{rs}+{}^3h_{rs}(\tau 
,\vec \sigma )=(1+{M\over r})
\delta_{rs}+o_4(r^{-3/2}),$\hfill\break
${}^3{\tilde \Pi}^{rs}(\tau ,\vec \sigma )\, {\rightarrow}
_{r\, \rightarrow \infty}\, {}^3k^{rs}(\tau ,\vec \sigma )=
o_3(r^{-5/2}),$\hfill\break
\hfill\break
$N(\tau ,\vec \sigma )= N_{(as)}(\tau ,\vec \sigma )
+n(\tau ,\vec \sigma ),\quad\quad n(\tau ,\vec 
\sigma )\, = O(r^{-(3+\epsilon )}),$\hfill\break
$N_{r}(\tau ,\vec \sigma )=N_{(as)r}(\tau ,\vec \sigma )+
n_{r}(\tau ,\vec \sigma ),\quad\quad 
n_{r}(\tau ,\vec \sigma )\, = O(r^{-\epsilon}),$\hfill\break
$N_{(as)A}(\tau ,\vec \sigma )\, {\buildrel {def} \over =}\,
(N_{(as)}\, ;\, N_{(as) r}\, )(\tau ,\vec \sigma )
=-{\tilde \lambda}_A(\tau )-{1\over 2}{\tilde \lambda}_{As}(\tau ) 
\sigma^{s},$\hfill\break
\hfill\break
$\Rightarrow \,\, {}^3e_{(a)r}(\tau ,\vec \sigma )=(1+{M\over {2r}}) \delta
_{(a)r} +o_4(r^{-3/2})$,\hfill\break
\hfill\break
with${}^3h_{rs}(\tau ,-\vec \sigma )={}^3h_{rs}(\tau ,\vec \sigma )$,
${}^3k^{rs}(\tau ,-\vec \sigma )=-{}^3k^{rs}(\tau ,\vec \sigma )$; here 
${}^3{\tilde \Pi}^{rs}(\tau ,\vec \sigma )$ is the momentum conjugate to the
3-metric ${}^3g_{rs}(\tau ,\vec \sigma )$ in ADM metric gravity.

These boundary conditions identify the class of spacetimes of Christodoulou and 
Klainermann\cite{ck} (they are near to Minkowski spacetime in a norm sense,
contain gravitational radiation but evade the singularity theorems, because 
they do not satisfy the hypothesis of conformal completion to get the
possibility to put control on the large time development of the solutions of
Einstein's equations). These spacetimes also satisfy the rest-frame condition 
$P^r_{ADM}=0$ (this requires ${\tilde \lambda}_{Ar}(\tau )=0$ like for Wigner
hyperplanes in parametrized Minkowski theories) and have vanishing shift 
functions (but non trivial lapse function).

After the addition of the surface term, the resulting canonical and Dirac
Hamiltonians of ADM metric gravity are\hfill\break
\hfill\break
${H}_{(c)ADM}=
\int d^3\sigma [(N_{(as)}+n) {\tilde {\cal H}}+(N_{(as)r}+
n_{r})\, {}^3{\tilde {\cal H}}^{r}](\tau ,\vec \sigma )
\mapsto$\hfill\break
$\mapsto {H}^{'}_{(c)ADM}=
\int d^3\sigma [(N_{(as)}+n) {\tilde {\cal H}}+(N_{(as)r}+
n_{r})\, {}^3{\tilde {\cal H}}^{r}](\tau ,\vec \sigma )+$\hfill\break
$+{\tilde \lambda}_A(\tau )
P^A_{ADM}+{\tilde \lambda}_{AB}(\tau ) J^{AB}_{ADM} =$\hfill\break
$=\int d^3\sigma [ n {\tilde {\cal H}}+n_{r}\, 
{}^3{\tilde {\cal H}}^{r}](\tau ,\vec \sigma )+
{\tilde \lambda}_A(\tau ) {\hat P}^A_{ADM}+{\tilde \lambda}_{AB}(\tau )
{\hat J}^{AB}_{ADM}\approx $\hfill\break
$\approx {\tilde \lambda}_A(\tau ) {\hat P}^A_{ADM}+{\tilde 
\lambda}_{AB}(\tau ) {\hat J}^{AB}_{ADM},$\hfill\break
\hfill\break
with the  ``weak conserved improper charges"
${\hat P}^A_{ADM}$, ${\hat J}_{ADM}^{AB}$ [they are volume integrals differing
from the weak charges by terms proportional to integrals of the constraints].
The previous splitting implies to replace the variables $N(\tau ,\vec \sigma )$,
$N_r(\tau ,\vec \sigma )$ with the ones ${\tilde \lambda}_A(\tau )$, ${\tilde 
\lambda}_{AB}(\tau )=-{\tilde \lambda}_{BA}(\tau )$, $n(\tau ,\vec \sigma )$,
$n_r(\tau ,\vec \sigma )$ [with conjugate momenta ${\tilde \pi}^A(\tau )$, 
${\tilde \pi}^{AB}(\tau )=-{\tilde \pi}^{BA}(\tau )$, ${\tilde \pi}^n(\tau
,\vec \sigma )$, ${\tilde \pi}^r_{\vec n}(\tau ,\vec \sigma )$]
in the ADM theory.

With these assumptions one has the following form of the line element (also its
form in tetrad gravity is given)\hfill\break
\hfill\break
$ds^2= \epsilon ( [N_{(as)}+n]^2 - [N_{(as) r}+n_r] {}^3e^r_{(a)}\, {}^3e^s
_{(a)} [N_{(as) s}+n_s] ) (d\tau )^2-$\hfill\break
$-2\epsilon [N_{(as) r}+n_r] d\tau d\sigma^r -\epsilon \, {}^3e_{(a)r}\, 
{}^3e_{(a)s} d\sigma^r d\sigma^s.$\hfill\break
\hfill\break
The final suggestion of Dirac is to modify ADM metric gravity in the
following way: \hfill\break
i) add the 10 new primary constraints $p^A_{(\infty )}-{\hat P}^A_{ADM}
\approx 0$, $J^{AB}_{(\infty )}-{\hat J}^{AB}_{ADM}\approx 0$, where
$p^A_{(\infty )}=b^A_{(\infty )(\mu )}p^{(\mu )}_{(\infty )}$, $J^{AB}_{(\infty
)}=b^A_{(\infty )(\mu )}b^B_{(\infty )(\nu )}S^{(\mu )(\nu )}_{(\infty )}$
[remember that $p^A_{(\infty )}$ and $J^{AB}_{(\infty )}$ satisfy a Poincar\'e 
algebra];\hfill\break
ii) consider ${\tilde \lambda}_A(\tau )$, ${\tilde \lambda}_{AB}(\tau )$, as
Dirac multipliers  for these 10
new primary constraints, and not as configurational 
(arbitrary gauge) variables coming from the
lapse and shift functions [so that there are no conjugate (vanishing) momenta 
${\tilde \pi}^A(\tau )$, ${\tilde \pi}^{AB}(\tau )$ and no associated Dirac
multipliers $\zeta_A(\tau )$, $\zeta_{AB}(\tau )$], in the
assumed Dirac Hamiltonian [it is finite and differentiable]\hfill\break
\hfill\break
$H_{(D)ADM}= \int d^3\sigma [ n {\tilde {\cal H}}+n_r {\tilde {\cal H}}^r
+\lambda_{n} {\tilde \pi}^n+\lambda^{\vec n}_r{\tilde \pi}^r_{\vec n}]
(\tau ,\vec \sigma )-$\hfill\break
$-{\tilde \lambda}_A(\tau ) [p^A_{(\infty )}-{\hat P}^A
_{ADM}]-{\tilde \lambda}_{AB}(\tau )[J^{AB}_{(\infty )}-
{\hat J}^{AB}_{ADM}]\approx 0,$\hfill\break
\hfill\break
The reduced phase space is still the ADM one: on the ADM variables there are 
only the secondary first class constraints ${\tilde {\cal H}}(\tau ,\vec 
\sigma )\approx 0$, ${\tilde {\cal H}}^r(\tau ,\vec \sigma )\approx 0$  
[generators of proper gauge transformations], because the other
first class constraints $p^A_{(\infty )}-{\hat P}^A_{ADM}\approx 0$, $J^{AB}
_{(\infty )}-{\hat J}^{AB}_{ADM}\approx 0$ do not generate improper gauge
transformations but eliminate 10 of the extra 20 variables.

In this modified ADM metric gravity, one has restricted the 3+1 splittings of 
$M^4$ to foliations whose leaves $\Sigma_{\tau}$ tend to Minkowski spacelike
hyperplanes asymptotically at spatial infinity in a direction independent
way. Therefore, these $\Sigma^{'}_{\tau}$ should be determined by the 10
degrees of freedom $x^{(\mu )}_{(\infty )}(\tau )$, $b^{(\mu )}_{(\infty 
)A}(\tau )$, like it happens for flat spacelike hyperplanes: this means that 
it must be possible to define a ``parallel transport" of the asymptotic tetrads 
$b^{(\mu )}_{(\infty )A}(\tau )$ to get well defined tetrads in each point of 
$\Sigma^{'}_{\tau }$. While it is not yet clear whether this can be done for 
${\tilde \lambda}_{AB}(\tau )\not= 0$, there is a solution for ${\tilde \lambda}
_{AB}(\tau )=0$. This case corresponds
to go to the Wigner-like hypersurfaces [the
analogue of the Minkowski Wigner hyperplanes with the asymptotic normal
$l^{(\mu )}_{(\infty )}=l^{(\mu )}_{(\infty )\Sigma}$ parallel to ${\hat P}
^{(\mu )}_{ADM}$]. Following the same procedure defined for Minkowski 
spacetime, one gets ${\bar S}^{rs}_{(\infty )}\equiv {\hat J}^{rs}_{ADM}$ [see 
Ref.\cite{lus1} for the definition of ${\bar S}^{AB}_{(\infty )}$],
${\tilde \lambda}_{AB}(\tau )=0$ and $-{\tilde \lambda}_A(\tau ) [p^A_{(\infty 
)}-{\hat P}^A_{ADM}] =-{\tilde \lambda}_{\tau}(\tau ) [\epsilon_{(\infty )}-
{\hat P}^{\tau}_{ADM}]+{\tilde \lambda}_{r}(\tau ) {\hat P}^{r}_{ADM}$
[$\epsilon_{(\infty )}=\sqrt{p^2_{(\infty )}}$], so that the final form of
these four surviving constraints is ($P^r_{ADM}=0$ implies ${\hat P}^r_{ADM}
\approx 0$; $M_{ADM}=\sqrt{{\hat P}^2_{ADM}}\approx {\hat P}^{\tau}_{ADM}$ is 
the ADM mass of the universe)\hfill\break
\hfill\break
$\epsilon_{(\infty )}-{\hat P}^{\tau}
_{ADM} \approx 0$, ${\hat P}^{r}_{ADM}\approx 0$.\hfill\break
\hfill\break
On this subclass of foliations [whose leaves $\Sigma^{(WSW)}_{\tau}$ will be 
called Wigner-Sen-Witten hypersurfaces; they define the intrinsic asymptotic 
rest frame of the gravitational field] one can introduce a parallel
transport by using the interpretation of Ref.\cite{p26} of the Witten
spinorial method of demonstrating the positivity of the ADM energy \cite{p27}.
Let us consider the Sen-Witten connection \cite{sen,p27} restricted to
$\Sigma_{\tau}^{(WSW)}$ (it depends on the trace of the extrinsic curvature
of $\Sigma_{\tau}^{(WSW)}$) and the spinorial Sen-Witten equation associated
with it. As shown in Ref.\cite{p28}, this spinorial equation can be
rephrased as an equation whose solution determines (in a surface dependent
dynamical way) a tetrad in each point of $\Sigma_{\tau}^{(WSW)}$ once it is 
given at spatial infinity (again this requires a direction independent
limit). Therefore, at spatial infinity there is a privileged congruence of
timelike observers, which replaces the concept of ``fixed stars" in the study
of the precessional effects of gravitomagnetism on gyroscopes and whose
connection with the definition of post-Newtonian coordinates has still to be
explored.

On the Wigner-Sen-Witten hypersurfaces 
the spatial indices have become spin-1 Wigner indices [they
transform with Wigner rotations under asymptotic Lorentz transformations].
As said for parametrized theories in Minkowski spacetime, in this special
gauge 3 degrees of freedom of the gravitational field 
[ an internal 3-center-of-mass variable ${\vec \sigma}_{ADM}[{}^3g,{}^3{\tilde 
\Pi}]$ inside the Wigner-Sen-Witten
hypersurface] become gauge variables, while ${\tilde x}^{(\mu )}_{(\infty )}$
[the canonical non covariant variable replacing $x^{(\mu )}_{(\infty )}$]
becomes a decoupled observer with his ``point particle clock"
\cite{ish,kuchar1} near spatial infinity.
Since the positivity theorems for the ADM 
energy imply that one has only timelike or lightlike
orbits of the asymptotic Poincar\'e group, the restriction to universes with
timelike ADM 4-momentum allows to define the M\o ller radius
$\rho_{AMD}=\sqrt{-{\hat W}^2_{ADM}}/{\hat P}^2_{ADM}$ from the asymptotic 
Poincar\'e Casimirs ${\hat P}^2_{ADM}$, ${\hat W}^2_{ADM}$.

By going from ${\tilde x}^{(\mu )}
_{(\infty )}$, $p^{(\mu )}_{(\infty )}$, to the canonical basis $T
_{(\infty )}=p_{(\infty )(\mu )}{\tilde x}^{(\mu )}_{(\infty )}/\epsilon
_{(\infty )}=p_{(\infty )(\mu )}x^{(\mu )}_{(\infty )}/\epsilon_{(\infty )}
{}{}{}$, $\epsilon_{(\infty )}$, $z^{(i)}_{(\infty )}=\epsilon_{(\infty )}
({\tilde x}^{(i)}_{(\infty )}-p^{(i)}_{(\infty )}{\tilde x}^{(o)}_{(\infty )}
/p^{(o)}_{(\infty )})$, $k^{(i)}_{(\infty )}=p^{(i)}_{(\infty )}/\epsilon
_{(\infty )}=u^{(i)}(p^{(\rho )}_{(\infty )})$, 
like in the flat case one finds that the final 
reduction requires the gauge-fixings $T_{(\infty )}-\tau \approx 0$ and
$\sigma^{r}_{ADM}\approx 0$, where $\sigma^{r}=\sigma^{r}_{ADM}$ is a variable 
representing the ``internal center of mass" of the 3-metric of the slice
$\Sigma_{\tau}$ of the asymptotically flat spacetime $M^4$. 
Since $\{ T_{(\infty )},\epsilon_{(\infty )} \}=-\epsilon$, with the
gauge fixing $T_{(\infty )}-\tau \approx 0$ one gets ${\tilde \lambda}_{\tau}
(\tau )\approx \epsilon$, and the final Dirac 
Hamiltonian is $H_D=M_{ADM}+{\tilde \lambda}_r(\tau ) {\hat P}^r_{ADM}$ with 
$M_{ADM}$ the natural physical Hamiltonian to 
reintroduce an evolution in the ``mathematical" $T_{(\infty )}\equiv
\tau$: namely in the rest-frame time identified with the parameter $\tau$
labelling the leaves $\Sigma_{\tau}^{(WSW)}$ of the foliation of $M^4$. 
Physical times (atomic clocks, ephemeridis time...) must be put in a local
1-1 correspondence with this ``mathematical" time. This point of view excludes
any Wheeler-DeWitt interpretation of an internal time (like the extrinsic York
one or the WKB times), which is used in closed universes of the
Einstein-Wheeler type.

All this construction holds also in our formulation of tetrad gravity 
(since it uses the ADM action) and in
its canonically reduced form in the 3-orthogonal gauges. The final physical 
Hamiltonian of tetrad gravity for the physical gravitational field 
is the reduced volume form of the ADM energy ${\hat P}^{\tau}_{ADM}[r_{\bar a}.
\pi_{\bar a}, \phi (r_{\bar a},\pi_{\bar a})]$ with the conformal factor $\phi$ 
solution of the reduced Lichnerowicz equation in the 3-orthogonal gauge with 
$\rho (\tau ,\vec \sigma )\approx 0$. The Hamilton-Dirac equations generated
by this Hamiltonian for $r_{\bar a}$, $\pi_{\bar a}$ generate the pair of
second order equations in normal form for $r_{\bar a}$ hidden in the Einstein
equations in this particular gauge.

Let us compare the standard generally covariant formulation of gravity based on
the Hilbert action with its invariance under $Diff\, M^4$ with the ADM
Hamiltonian formulation.

Regarding the 10 Einstein equations of the standard approach,
the Bianchi identities imply that four equations are linearly dependent on the 
other six ones and their gradients. Moreover, the four combinations of
Einstein's equations projectable to phase space (where they become the
secondary first class superhamitonian and supermomentum constraints of canonical
metric gravity) are independent from the accelerations being restrictions
on the Cauchy data. As a consequence
the Einstein equations have solutions, in which the ten
components ${}^4g_{\mu\nu}$ of the 4-metric depend on only two truly dynamical
degrees of freedom (defining the physical gravitational field) and on eight
undetermined degrees of freedom.
This transition from the ten components ${}^4g
_{\mu\nu}$ of the tensor ${}^4g$ in some atlas of $M^4$ to the 2 
(deterministic)+8 (undetermined) degrees of freedom breaks general covariance,
because these quantities are neither tensors nor invariants under
diffeomorphisms (their functional form is atlas dependent).

Since the Hilbert action is invariant under $Diff\, M^4$, one usually says
that a ``dynamical
gravitational field" is a 4-geometry over $M^4$, namely an equivalence
class of spacetimes $(M^4, {}^4g)$, solution of Einstein's equations, modulo
$Diff\, M^4$. See, however, the interpretational problems about what is 
observable in general relativity for instance in Refs.\cite{rove,rov}, in
particular the facts that at least before the restriction to the solutions
of Einstein's equations
i) scalars under $Diff\, M^4$, like ${}^4R$, are not
Dirac's observables but gauge dependent quantities; ii) the functional form
of ${}^4g_{\mu\nu}$ in terms of the physical gravitational field and, 
therefore, the angle and distance properties of material bodies and the
standard procedures of defining measures of length and time based on the
line element $ds^2$, are gauge dependent.

Instead in the ADM formalism with the extra notion of 3+1 splittings of $M^4$,
the (tetrad) metric ADM action (differing from the Hilbert one by a surface
term) is quasi-invariant under the (14) 8 types of gauge transformations
which are the pull-back of the Hamiltonian group ${\cal G}$ of gauge
transformations, whose generators are the first class constraints of the theory
. The Hamiltonian group ${\cal G}$ has a subgroup (whose generators are the 
supermomentum and superhamiltonian constraints) formed by the diffeomorphisms
of $M^4$ adapted to its 3+1 splittings, $Diff\, M^{3+1}$ [it is different
from $Diff\, M^4$]. Moreover, the Poisson algebra of the supermomentum and 
superhamiltonian constraints reflects the embeddability in $M^4$ of the
foliation associated with the 3+1 splitting \cite{tei}.

Now in tetrad gravity
the interpretation of the 14 gauge transformations and of their gauge fixings
(it is independent from the presence of matter) is the 
following [a tetrad in a point of $\Sigma_{\tau}$ is a local observer]
:\hfill\break
i) the gauge fixings of the gauge boost parameters associated with the 3 boost
constraints and of the gauge angles associated with the 3 rotation constraints
are equivalent to choose the congruence of timelike observers to be used as a
standard of non rotation;\hfill\break
ii) the gauge fixings of the 3 gauge parameters associated with the passive
space diffeomorphisms [$Diff\, \Sigma_{\tau}$; change of coordinates charts]
are equivalent to a fixation of 3 standards of length by means of a choice of
a coordinate system on $\Sigma_{\tau}$ [the measuring apparatus (the ``rods")
should be defined in terms of Dirac's observables for some kind of matter, after
its introduction into the theory];\hfill\break
iii) according to constraint theory the choice of 3-coordinates on $\Sigma
_{\tau}$ induces the gauge fixings of the 3 shift functions [i.e. of ${}^4g
_{oi}$], whose gauge nature is connected with the ``conventionality of
simultaneity" \cite{havas} [therefore, the gauge fixings are equivalent
to a choice of synchrinization of clocks 
and, as a consequence, to a statement about the
isotropy or anisotropy of the velocity of light in that gauge];\hfill\break
iv) the gauge fixing on the the momentum $\rho (\tau ,\vec \sigma )$ conjugate
to the conformal factor of the 3-metric [this gauge variable is the source of 
the gauge dependence of 4-tensors and of the scalars under $Diff\, M^4$, 
together with the gradients of the lapse and shift functions] is a nonlocal 
statement about the extrinsic curvature of the leaves $\Sigma_{\tau}$ of the 
given 3+1 splitting of $M^4$; since the superhamiltonian constraint produces 
normal deformations of $\Sigma_{\tau}$ \cite{tei} and, therefore, transforms a 
3+1 splitting of $M^4$ into another one (the ADM formulation is independent from
the choice of the 3+1 splitting), this gauge fixing is equivalent to the
choice of a particular 3+1 splitting;\hfill\break
v) the previous gauge fixing induces the gauge fixing of the lapse function
(which determines the packing of the leaves $\Sigma_{\tau}$ in the chosen
3+1 splitting) and, therefore, is equivalent to the fixation of a standard of
proper time [again ``clocks" should be built with the Dirac's observables
of some kind of matter].

In the Hamiltonian formalism it is natural to define
a ``Hamiltonian kinematical gravitational field" as the equivalence
class of spacetimes modulo the Hamiltonian group ${\cal G}$, and different 
members of the equivalence class have in general different 4-Riemann tensors
[these equivalence classes are connected with the conformal 3-geometries of the
Lichnerowicz-York approach and contain different gauge-related 4-geometries].
Then, a ``Hamiltonian dynamical gravitational field" is defined as a
Hamiltonian kinematical gravitational fields which is solution of the
Hamilton-Dirac equations generated by the weak ADM energy ${\hat P}^{\tau}
_{ADM}$. Since the Hilbert and ADM actions, even if they have different local 
symmetries and invariances, both generate the same Einstein equations, the 
equivalence classes of the ``Hamiltonian dynamical gravitational fields" and of 
the standard ``dynamical gravitational fields" (a 4-geometry solution of 
Einstein's equations) coincide. Indeed,
on the solutions of Einstein's equations the gauge transformations generated by 
the superhamiltonian constraint (normal deformations of $\Sigma_{\tau}$) and
those generated by the canonical momenta of the lapse and shift functions 
together with the $\Sigma_{\tau}$ diffeomorphisms generated by the supermomentum
constraints are restricted by the Jacobi equations associated to Einstein's 
equations to be those Noether symmetries of the ADM action which are also 
dynamical symmetries of the Hamilton equations and therefore they are a subset
of the spacetime diffeomorphisms $Diff\, M^4$ (all of which are dynamical 
symmetries of Einstein's equations).

The 3-orthogonal gauges of tetrad gravity are the equivalent of the Coulomb 
gauge in classical electrodynamics (like the harmonic gauge is the equivalent 
of the Lorentz gauge). Only after a complete gauge fixing the 4-tensors and 
the scalars under $Diff\, M^4$ become measurable quantities (like the 
electromagnetic vector potential in the Coulomb gauge):  an experimental
laboratory does correspond by definition to a completely fixed gauge.  At this 
stage it becomes acceptable the proposal of Komar\cite{komar} and
Bergmann\cite{be} of identifying the points
of a spacetime $(M^4, {}^4g)$, solution of the Einstein's equations in
absence of matter, in a way invariant under spacetime diffeomorphisms, by using 
four bilinears and trilinears in the Weyl tensors, scalar under $Diff\, M^4$
and called ``individuating fields''(see also Refs.\cite{rove,rov}),
which do not depend on the lapse and shift functions (but only on the gauge
variables corresponding to the 3-coordinates on $\Sigma_{\tau}$ and to the
momentum conjugate to the conformal factor of the 3-metric, so that these
fields carry the information on the choice of the 3-coordinates and of a
generalized extrinsic time), to build 
``physical 4-coordinates" (in each completely fixed gauge they depend only on 
the two canonical pairs of Dirac's observables of the gravitational field),
justifying a posteriori the standard measurement
theory presented in all textbooks on general relativity, which presuppones the
individuation of spacetime points.

Our approach breaks the general covariance of general relativity completely by
going to the special 3-orthogonal gauges. But this is done in a way
naturally associated with theories with first class
constraints: the global
Shanmugadhasan canonical transformations (when they exist) 
correspond to privileged Darboux charts for presymplectic manifolds defined
by the first class constraints. Therefore, the gauges identified by 
these canonical transformations should have a special (till now unexplored) 
role also also in generally covariant theories, in which traditionally one 
looks for observables invariant under diffeomorphisms
 and not for not generally covariant 
Dirac observables.

Let us remember that Bergmann\cite{be} made the following critique of 
general covariance: it would be desirable to restrict the group of
coordinate transformations (spacetime diffeomorphisms) in such a way that it
could contain an invariant subgroup describing the coordinate transformations 
that change the frame of reference of an outside observer (these 
transformations could be called Lorentz transformations; see also the
comments in Ref.\cite{ll} on the asymptotic behaviour of coordinate
transformations); the remaining 
coordinate transformations would be like the gauge transformations of
electromagnetism. This is what we have done. In this way
``preferred' coordinate systems will emerge (the WSW hypersurfaces with their
preferred congruences of timelike observers whose 4-velocity becomes
asymptotically normal to $\Sigma_{\tau}^{(WSW)}$ at spatial infinity), 
which, as said by Bergmann, are
not ``flat": while the inertial coordinates are determined experimentally by 
the observation of trajectories of force-free bodies, these intrinsic 
coordinates can be determined only by much more elaborate experiments
(probably with gyroscopes), since
they depend, at least, on the inhomogeneities of the ambient gravitational 
fields.
See also Ref.\cite{ellis} for other critics to general covariance: very often
to get physical results one uses preferred coordinates not merely for
calculational convenience, but also for understanding (this
fact has been formalized as the ``principle of restricted covariance").

Since in the 3-orthogonal gauges we have the physical canonical basis $r_{\bar
a}$, $\pi_{\bar a}$, it is possible, but only in absence of matter, to define 
``void spacetimes" as the equivalence class of spacetimes ``without 
gravitational field", whose members in the 3-orthogonal gauges are obtained 
by adding by hand the second class constraints $r_{\bar a}(\tau ,\vec 
\sigma )\approx 0$, $\pi_{\bar a}(\tau ,\vec \sigma )\approx 0$ [one gets $\phi
(\tau ,\vec \sigma )=1$ as the relevant solution of the reduced Lichnerowicz
equation] and, in particular, their Poincar\'e charges  vanish (this corresponds
to the exceptional $p^{(\mu )}=0$ orbit of the Poincar\'e group and shows the
peculiarity of these solutions with zero ADM mass). It is expected that the
void spacetimes can be defined in a gauge-independent way by adding to the ADM 
action the requirement that the leaves $\Sigma_{\tau}$ of the 3+1 splitting be
3-conformally flat, namely that the Cotton-York 3-conformal tensor vanishes.
The members of this equivalence class (the extension to general 
relativity of the Galilean non inertial coordinate systems with their Newtonian 
inertial forces) are gauge equivalent to Minkowski spacetime with 
Cartesian coordinates and it is expected that they describe pure acceleration
effects without physical gravitational field (no tidal effects).

See Ref.\cite{pl} for the $c\rightarrow \infty$ contraction of the ADM action of
metric gravity: a theory with 26 independent fields (most of them describe 
inertial forces) and with general Galileo covariance has been obtained. This
formulation of Newton gravity should be the natural nonrelativistic limit of
Einstein's general relativity in the framework of singular Lagrangians;
however, its connection with the post-Newtonian approximations has still to be 
explored.

If we add \cite{russo4}
to the tetrad ADM action the action for N scalar particles with
positive energy in the form of Ref.\cite{lus1} [where it was given on
arbitrary Minkowski spacelike hypersurfaces], the only constraints which are
modified are the superhamiltonian one, which gets a dependence on the matter 
energy density ${\cal M}(\tau ,\vec \sigma )$, and the 3 space diffeomorphism
ones, which get a dependence on the matter momentum density ${\cal M}_r(\tau
,\vec \sigma )$. The canonical reduction and the determination of the Dirac 
observables can be done like in absence of matter. However, the reduced 
Lichnerowicz equation for the conformal factor of the 3-metric in the 
3-orthogonal gauge and with $\rho (\tau ,\vec \sigma )\approx 0$ acquires now
an extra dependence on ${\cal M}(\tau ,\vec \sigma )$ and ${\cal M}_r(\tau 
,\vec \sigma )$. 

Since, as a preliminary result, we are interested in
identifying explicitly the instantaneous action-at-a-distance 
(Newton-like and gravitomagnetic) potentials \hfill\break
among particles hidden in tetrad
gravity (like the Coulomb potential is hidden in the electromagnetic gauge
potential), we shall make the strong approximation of neglecting the (tidal)
effects of the physical gravitational field by putting $r_{\bar a}
(\tau ,\vec \sigma )\approx 0$, $\pi_{\bar a}(\tau ,\vec \sigma )\approx 0$,
even if it is not strictly consistent with the Hamilton-Dirac equation
(extremely weak gravitational fields).
If, furthermore, we develop the conformal factor $\phi (\tau ,\vec 
\sigma )$ in a formal series in the Newton constant G [$\phi =1+\sum_{n=1}
^{\infty} G^n \phi_n$], one can find a solution $\phi = 1+G \phi_1$ at order G
(post-Minkowskian approximation) of the reduced Lichnerowicz equation
where we put $r_{\bar a}=\pi_{\bar a}=0$. However, 
due to a self-energy divergence in $\phi$ evaluated at the positions ${\vec
\eta}_i(\tau )$ of the particles, one needs to rescale the bare masses to
physical ones, $m_i\, \mapsto \, \phi^{-2}(\tau ,{\vec \eta}_i(\tau )) 
m_i^{(phys)}$, and to make a regularization of the type defined in Refs.
\cite{ein}. Then, the regularized solution for $\phi$ can be put in the reduced 
form of the ADM energy, which becomes [${\vec \kappa}_i(\tau )$ are the
particle momenta conjugate to ${\vec \eta}_i(\tau )$; ${\vec n}_{ij}=
[{\vec \eta}_i-{\vec \eta}_j]/|{\vec \eta}_i-{\vec \eta}_j|$]\hfill\break
\hfill\break
${\hat P}^{\tau}_{ADM}=\sum_{i=1}^Nc\sqrt{m_i^{(phys) 2}c^2+{\vec \kappa}^2
_i(\tau )}-$\hfill\break
$-{G\over {c^2}}\sum_{i\not= j} {{ \sqrt{m_i^{(phys) 2}c^2+{\vec 
\kappa}^2_i(\tau )}\, \sqrt{m_j^{(phys) 2}c^2+{\vec \kappa}^2_j(\tau )} }\over
{|{\vec \eta}_i(\tau )-{\vec \eta}_j(\tau )|}}-$\hfill\break
$-{G\over {8c^2}} \sum_{i\not= j} {{ 3{\vec \kappa}_i(\tau )\cdot {\vec 
\kappa}_j(\tau )-5 {\vec \kappa}_i(\tau )\cdot {\vec n}_{ij}(\tau ) 
{\vec \kappa}_j(\tau )\cdot {\vec n}_{ij}(\tau )}\over
{|{\vec \eta}_i(\tau )-{\vec \eta}_j(\tau )|}} + O(G^2,r_{\bar a},\pi_{\bar a})
.$\hfill\break
\hfill\break
One sees the Newton-like and the gravitomagnetic (in the sense of York)
potentials (both of them need regularization) at the post-Minkowskian level
(order G but exact in c) emerging from the tetrad ADM version of Einstein
general relativity when we ignore the tidal effects. 
For G=0 we recover N free scalar particles on the Wigner
hyperplane in Minkowski spacetime, as required by deparametrization. For
$c\, \rightarrow \, \infty$, we get the post-Newtonian Hamiltonian\hfill\break
\hfill\break
$H_{PN}=\sum_{i=1}^N{{ {\vec \kappa}_i^2(\tau )}\over {2m_i^{(phys)}}}
(1-{{2G}\over {c^2}}\sum_{j\not= i}{{m_j^{(phys)}}\over 
{|{\vec \eta}_i(\tau )-{\vec \eta}_j(\tau )|}})-{G\over 2}\sum_{i\not= j}
{{m_i^{(phys)}\, m_j^{(phys)}}\over
{|{\vec \eta}_i(\tau )-{\vec \eta}_j(\tau )|}}-$\hfill\break
$-{G\over {8c^2}}\sum_{i\not= j} {{ 3{\vec \kappa}_i(\tau )\cdot {\vec 
\kappa}_j(\tau )-5 {\vec \kappa}_i(\tau )\cdot {\vec n}_{ij}(\tau ) 
{\vec \kappa}_j(\tau )\cdot {\vec n}_{ij}(\tau )}\over
{|{\vec \eta}_i(\tau )-{\vec \eta}_j(\tau )|}} + O(G^2,r_{\bar a},\pi_{\bar a}),
$\hfill\break
\hfill\break
which is of the type of the ones implied by the results of
Refs.\cite{ein,droste} [the differences 
are probably connected with the use of different coordinate
systems and with the fact that one has essential singularities on the particle
worldlines and the need of regularization].

The main open problems now under investigation are: i)
the linearization of the theory in the 3-orthogonal gauges in presence of
matter to find the 3-orthogonal gauge description of gravitational waves
and to go beyond the previous instantaneous post-Minkowskian approximation at
least in the 2-body case relevant for the motion of binaries; ii)
the replacement of scalar particles with spinning ones to identify the
precessional effects (like the Lense-Thirring one) of gravitomagnetism;
iii) the coupling to perfect fluids for the simulation of rotating stars and
for the comparison with the post-Newtonian approximations; iv)
the coupling of  tetrad gravity to the
electromagnetic field, to fermion fields and then to the standard model,
trying to make to reduction to Dirac's observables in all these cases and to
study their post-Minkowskian approximations;
v) the 
quantization of tetrad gravity in the 3-orthogonal gauge with $\rho (\tau 
,\vec \sigma )\approx 0$ (namely after a complete breaking of general 
covariance): for each perturbative (in G) solution of the reduced 
Lichnerowicz equation one defines a Schroedinger equation in $\tau$ for a wave 
functional $\Psi [\tau ; r_{\bar a}]$ with the associated quantized ADM energy
${\hat P}^{\tau}_{ADM}[r_{\bar a}, i{{\delta}\over {\delta r_{\bar a}}} ]$
as Hamiltonian; no problem of physical scalar product is present, but only
ordering problems in the Hamiltonian; moreover, one has the M\o ller radius as
a ultraviolet cutoff. Also a comparison with ``loop quantum gravity"
\cite{ars}, which respects general covariance but only for fixed lapse
and shift functions,  has still to be done.

Therefore, a well defined classical stage for a unified description of the four
interactions is emerging, even if many aspects have only been clarified at a
heuristic level so that a big effort from both mathematical and theoretical
physicists is still needed. It will be exciting to see whether in the next
years some reasonable quantization picture will develop from this classical 
framework.

\end{document}